\documentclass[ArXiv,AXISWP,accept,moreauthors,pdftex,10pt,letterpaper]{Definitions/axis} 

\firstpage{1} 
\pubvolume{xx}
\issuenum{1}
\articlenumber{5}
\pubyear{2023}
\copyrightyear{2023}

\usepackage{comment}
\usepackage{sidecap}
\sidecaptionvpos{figure}{t}
\usepackage{amsmath,amssymb}

\newcommand{\Mpc}{\rm\; Mpc}
\newcommand{\kpc}{\rm\; kpc}
\newcommand{\pc}{\rm\; pc}
\newcommand{\km}{\rm\; km}

\newcommand{\cm}{\rm\; cm}

\newcommand{\cmsq}{\hbox{$\cm^2\,$}}

\newcommand{\yr}{\rm\; yr}

\newcommand{\Myr}{\rm\; Myr}
\newcommand{\s}{\rm\; s}

\newcommand{\ks}{\rm\; ks}

\newcommand{\Ms}{\rm\; Ms}
\newcommand{\K}{\rm\; K}
\newcommand{\Msun}{\hbox{$\rm\thinspace M_{\odot}$}}

\newcommand{\Msunpyr}{\hbox{$\Msun\yr^{-1}\,$}}
\newcommand{\keV}{\rm\; keV}

\newcommand{\erg}{\rm\; erg}

\newcommand{\ergpcmsqps}{\hbox{$\erg\cm^{-2}\s^{-1}\,$}}

\newcommand{\ergps}{\hbox{$\erg\s^{-1}\,$}}

\newcommand{\ctspasecsq}{\hbox{$\rm\thinspace counts\pasecsq$}}

\newcommand{\kmps}{\hbox{$\km\s^{-1}\,$}}

\newcommand{\amin}{\rm\; arcmin}

\newcommand{\asec}{\rm\; arcsec}

\newcommand{\pcmsq}{\hbox{$\cm^{-2}\,$}}

\newcommand{\pasecsq}{\hbox{$\asec^{-2}\,$}}
\Title{The evolution of galaxies and clusters at high spatial resolution with AXIS}

\Author{H.~R. Russell$^{1}$, L.~A. Lopez$^{2}$, 
   S.~W. Allen$^{3,4,5}$,
  G. Chartas$^{6}$,
  P.~P. Choudhury$^{7,8}$,
  R.~A. Dupke$^{9,10}$,
  A.~C. Fabian$^7$,
  A.~M. Flores$^{3,4}$,
  K. Garofali$^{11}$,
  E. Hodges-Kluck$^{11}$,
  M.~J. Koss$^{12}$,
  L. Lanz$^{13}$,
  B.~D. Lehmer$^{14}$,
  J.-T. Li$^{15}$,
  W.~P. Maksym$^{16}$,
  A.~B. Mantz$^{4}$,
  M. McDonald$^{17,18}$,
  E.~D. Miller$^{18}$, 
  R.~F. Mushotzky$^{19}$,
  Y. Qiu$^{20}$,
  C.~S. Reynolds$^{19,21}$,
  F. Tombesi$^{22}$,
  P. Tozzi$^{23}$,
  A. Trindade-Falc\~{a}o$^{16}$,
  S.~A. Walker$^{24}$,
  K.-W. Wong$^{25}$,
  M. Yukita$^{11}$ and
  C. Zhang$^{26}$
}

\AuthorNames{H.~R. Russell, L.~A. Lopez, S.~W. Allen, G. Chartas, P.~P. Choudhury, R.~A. Dupke, A.~C. Fabian, A.~M. Flores, K. Garofali, E. Hodges-Kluck, M.~J. Koss, L. Lanz, B.~D. Lehmer, J.-T. Li, W.~P. Maksym, A.~B. Mantz, M. McDonald, E.~D. Miller, R.~F. Mushotzky, Y. Qiu, C.~S. Reynolds, F. Tombesi, P. Tozzi, A. Trindade-Falc\~{a}o, S.~A. Walker, K.~W. Wong, M. Yukita and C. Zhang}

\address{%
$^{1}$ \quad School of Physics \& Astronomy, University of Nottingham, University Park, Nottingham NG7 2RD, UK\\
$^{2}$ \quad Department of Astronomy, The Ohio State University, 140 W. 18th Ave., Columbus, OH 43210, USA\\
$^{3}$ \quad Department of Physics, Stanford University, 382 Via Pueblo Mall, Stanford, CA 94305, USA\\
$^{4}$ \quad Kavli Institute for Particle Astrophysics and Cosmology, Stanford University, 452 Lomita Mall, Stanford, CA\\ \hspace{4.7mm} 94305, USA\\
$^{5}$ \quad SLAC National Accelerator Laboratory, 2575 Sand Hill Road, Menlo Park, CA 94025, USA\\
$^{6}$ \quad Department of Physics and Astronomy, College of Charleston, Charleston, SC, 29424, USA\\
$^7$ \quad Institute of Astronomy, University of Cambridge, Madingley Road, Cambridge CB3 0HA, UK\\
$^8$ \quad Clarendon Laboratory, University of Oxford, Parks Rd, Oxford OX13PU, UK\\
$^{9}$ \quad University of Michigan, Department of Astronomy, 1085 South University Ave., Ann Arbor, MI, USA\\
$^{10}$ \quad Observat\'orio Nacional,  Rua Gal. Jos\'e Cristino 77, S\~ao Crist\'ov\~ao, Rio de Janeiro\\
$^{11}$ \quad NASA/GSFC, Greenbelt, MD 20771, USA\\
$^{12}$ \quad Eureka Scientific, 2452 Delmer Street Suite 100, Oakland, CA 94602-3017, USA\\
$^{13}$ \quad Department of Physics, The College of New Jersey, Ewing, NJ 08628, USA\\
$^{14}$ \quad Department of Physics, University of Arkansas, Fayetteville, AR 72701, USA\\
$^{15}$ \quad Purple Mountain Observatory, Chinese Academy of Sciences, 10 Yuanhua Road, Nanjing 210023, People’s\\ \hspace{5.7mm} Republic of China\\
$^{16}$ \quad Center for Astrophysics | Harvard \& Smithsonian, 60 Garden St., Cambridge, MA 02138, USA\\
$^{17}$ \quad Department of Physics, Massachusetts Institute of Technology, Cambridge, MA 02139, USA\\
$^{18}$ \quad Kavli Institute for Astrophysics and Space Research, Massachusetts Institute of Technology, Cambridge, MA\\ \hspace{5.7mm} 02139, USA\\
$^{19}$ \quad Department of Astronomy, University of Maryland, College Park, MD 20742-2421, USA\\
$^{20}$ \quad Kavli Institute for Astronomy and Astrophysics, Peking University, Beijing 100871, China\\
$^{21}$ \quad Joint Space-Science Institute (JSI), College Park, MD 20742-2421, USA\\
$^{22}$ \quad Dipartimento di Fisica, Universit\`{a} degli Studi di Roma “Tor Vergata”, Via della Ricerca Scientiﬁca 1, 00133\\ \hspace{5.7mm} Roma, Italy\\
$^{23}$ \quad INAF – Osservatorio Astroﬁsico di Arcetri, Largo E. Fermi, 50122 Firenze, Italy\\
$^{24}$ \quad Department of Physics and Astronomy, University of Alabama in Huntsville, Huntsville, AL 35899, USA\\
$^{25}$ \quad Department of Physics, SUNY Brockport, Brockport, NY, 14420, USA\\
$^{26}$ \quad Department of Astronomy and Astrophysics, The University of Chicago, Chicago, IL 60637, USA\\
}

\abstract{Stellar and black hole feedback heat and disperse surrounding cold gas clouds, launching gas flows off circumnuclear and galactic disks and producing a dynamic interstellar medium.  On large scales bordering the cosmic web, feedback drives enriched gas out of galaxies and groups, seeding the intergalactic medium with heavy elements.  In this way, feedback shapes galaxy evolution by shutting down star formation and ultimately curtailing the growth of structure after the peak at redshift $2-3$.  To understand the complex interplay between gravity and feedback, we must resolve both the key physics within galaxies and map the impact of these processes over large scales, out into the cosmic web.  The Advanced X-ray Imaging Satellite (AXIS) is a proposed X-ray probe mission for the 2030s with arcsecond spatial resolution, large effective area, and low background.  AXIS will untangle the interactions of winds, radiation, jets, and supernovae with the surrounding ISM across the wide range of mass scales and large volumes driving galaxy evolution and trace the establishment of feedback back to the main event at cosmic noon.  \emph{This White Paper is part of a series commissioned for the AXIS Probe mission concept; additional AXIS White Papers can be found at the  \href{http://axis.astro.umd.edu/}{AXIS website} with a mission overview \href{https://arxiv.org/abs/2311.00780}{here}}.}

\begin{document}

%%%%%%%%%%%%%%%%%%%%%%%%%%%%%%%%%%%%%%%%%%

\tableofcontents
\listoffigures

\section{Introduction}  

\noindent Galaxies form and evolve through a vigorous cosmic struggle.  The inexorable pull of gravity drives the assembly of galaxies by drawing primordial gas into dark matter halos and collapsing cold clouds into stars.  At most twenty percent of baryons are converted into stars [e.g. \citealp{Wechsler18}].  This inexplicable inefficiency across the galaxy mass scale is the result of high-energy processes around stars and supermassive black holes.  Stellar feedback, via stellar winds and supernovae, heats and disperses cold gas clouds and collectively inflates large gas bubbles that lift hot gas and metals into galactic halos [for reviews, see \citealp{Veilleux05,Heckman17}].  In massive galaxies, more powerful feedback from active galactic nuclei (AGN), in the form of radiation, winds and relativistic jets, drives large outflows, suppresses cooling of hot atmospheres, and produces a correlative slowing of black hole activity \cite{Fabian12}.  Feedback indelibly shapes galaxy evolution on all scales of the galaxy mass function, back to the peak of galaxy growth at $z\sim2$ and beyond.  

X-ray observations reveal the key interactions on all spatial scales: the acceleration of winds and jets close to black holes, the impact zones where stellar winds, supernovae, AGN winds, and jets meet the ISM, and the distribution of energy and metal-rich gas across galaxy halos.  Arcsecond spatial resolution is essential for disentangling the complex interactions of winds, radiation, jets, and supernovae with the surrounding gas.  However, only when combined with high throughput can the key scales of individual HII regions, star clusters, wind shocks, and jet-blown bubbles be reached across the galaxy population, tracing the establishment of feedback back to the main event at cosmic noon.  AXIS now capitalizes on the key technological advance of precision-focusing with light-weight X-ray mirrors to deliver $1.5\asec$ spatial resolution on-axis, $1.6\asec$ on average across the $24\amin$ field of view, and a factor of $\sim10$ increase in collecting area over existing facilities \cite{Reynolds23}.  AXIS will answer major questions on feedback including:

\begin{itemize}
\item{} How do star-forming structures shape the ISM?  What are the conditions for breakout or outflow?

\item{} How do AGN winds and jets inject energy and drive outflows in individual galaxies?

\item{} How and when did AGN feedback begin to shape the evolution of galaxy clusters and groups?

\item{} How and when did the hot and diffuse universe become enriched with heavy elements?
\end{itemize}

\noindent Furthermore, in combination with the low particle background of a low-inclination low Earth orbit, AXIS will unveil the faint and diffuse X-ray universe: the remarkable interconnected gas flows that comprise the cosmic web, giant filaments that feed cluster assembly, and the missing baryons in the circumgalactic medium [CGM, see e.g., \citealp{Walker_19_whitepaper}].

In this white paper, we discuss how AXIS will transform our understanding of all aspects of feedback in galaxies, a main science driver of the mission, and we present the far wider impact that AXIS will precipitate across all areas of galaxy formation and evolution, which will be realized in the $>70\%$ of general observer time.

\section{Stellar Feedback in Nearby Galaxies}
\vspace{-0.5\baselineskip}
\noindent (Contributed by L. A. Lopez, M. Yukita, E. Hodges-Kluck and K. Garofali)
\vspace{0.5\baselineskip}

\begin{figure}
    \centering
    \includegraphics[width=0.9\columnwidth]{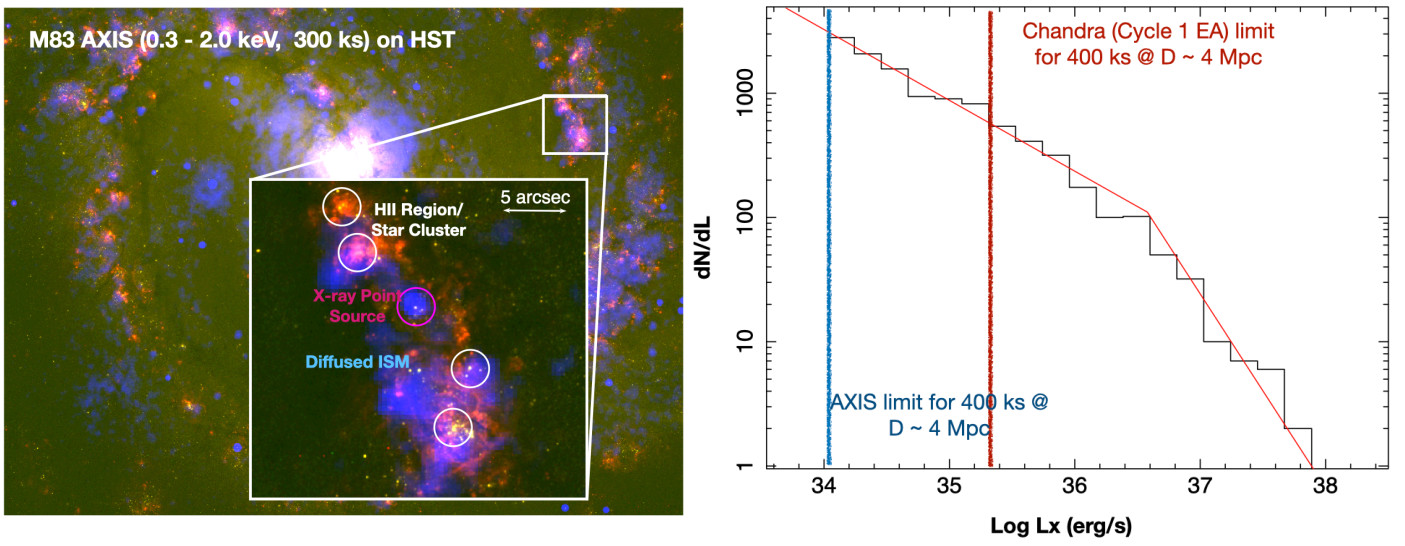}
    \caption[AXIS simulation of stellar feedback in M83]{AXIS $300\ks$ simulation of the soft-band ($0.3-2\keV$) image of the spiral galaxy M83. X-ray binaries and diffuse hot gas are apparent, occupying the star-forming regions and inflating bubbles through the collective influence of supernovae and stellar winds. With AXIS sensitivity and spatial resolution, spectro-imaging analyses across arcsecond ($10-30\pc$) scales are possible. Right: A sharp, sensitive survey will definitively measure the luminosity function of $>5000$ star clusters down to below the ``knee" in the H$\alpha$ luminosity function. \textit{Chandra} is not able to achieve these depths in multiple Ms even with capabilities at launch.}
    \label{fig:m83}
\end{figure}

\noindent 
Over the past decade, observational and theoretical investigations have revealed that the large-scale behavior of galaxies (e.g., the shape and structure of galaxies, the rate and efficiency of star formation) emerge from local phenomena, particularly stellar feedback \cite{hopkins14}. The injection of energy and momentum by stars has a profound influence on their surroundings. For example, stellar winds and supernovae from massive star clusters inflate bubbles that disrupt natal clouds \cite{chevance20}, triggering star formation on their periphery and dispersing gas and metals that become fodder for the next generation of stars \citep{krumholz19}. Collections of these bubbles build on each other, and in the extreme cases of starburst galaxies, form superheated gas that can drive galactic winds that eject much of the ISM gas into galactic halos \cite{Veilleux05}.

%Our understanding of the small-scale processes of feedback is critical as this is the physical scale (10–200 pc scales of H{\sc ii} regions and giant molecular clouds) on which feedback operates. X-ray provides essential information regarding stellar feedback driven by hot gas, as X-ray carries most of the energy. However, our grasp of the stellar feedback traced from X-ray emission on these scales relies mostly on resolved studies of the Milky Way and Magellanic Clouds. X-rays are the missing piece in our high-resolution, multiwavelength view of galaxies, and it is critical to explore a diverse set of environments/galaxies.

Stellar feedback is often cited as one of the biggest uncertainties in star and galaxy formation [e.g., \citealp{dobbs14,hopkins14}], stemming from the need to constrain how the energy and momentum of star clusters is partitioned and couples to the ISM. Feedback operates through several channels (e.g., photoionization, radiation, stellar winds, supernovae, cosmic rays), and their effectiveness and impact vary, depending on star cluster mass, age, metallicity, and environment. An accurate energetics inventory that measures feedback efficiency is crucial to understand galactic ecosystems. This endeavor necessitates a multiwavelength perspective to connect cluster stars (UV, optical, and infrared) to HII regions (X-ray, UV, and optical) and the surrounding ISM (mm, radio, and infrared). 

X-ray observations are a critical missing piece and are essential to quantitatively measure the energetics because they trace the internal energy remaining in the hot gas.  How and where this energy is lost remains an open question [e.g., \citealp{rosen14,lancaster21}]. However, HII regions have sizes of $1-200\pc$ and typical luminosities of $L_{\rm X} = 10^{32} - 10^{37}\ergps$, largely inaccessible by modern X-ray facilities even in the Local Group. To date, only the brightest ($>10^{37}\ergps$) tens of sources in the Local Group have been studied because \textit{Chandra} lacks the sensitivity to detect fainter sources and \textit{XMM-Newton} cannot resolve them spatially. 

%AXIS has the angular resolution, sensitivity, and energy resolution necessary to measure the X-ray luminosities down to $L_{\rm X} > 10^{34}~{\rm erg}~{\rm s}^{-1}$, temperatures, number densities, column densities, and metallicities in a representative sample of HII regions in the Local Universe. Simultaneously, AXIS will constrain the non-thermal luminosities from cosmic ray electrons and X-ray binaries. 

AXIS's high spatial resolution and large, soft X-ray collecting area will enable the first representative inventory of the X-ray properties of individual star-forming regions ($1-10\asec$ scales) in nearby ($<7\Mpc$) galaxies (e.g., Fig. \ref{fig:m83}). AXIS will measure the X-ray signatures (e.g., temperature, density, luminosity) of $>5000$ star clusters down to $L_{\rm X}\sim10^{34}\ergps$ in these galaxies, sufficient to detect the X-ray counterparts of star cluster populations and to produce the first-ever X-ray luminosity function of this population. Nearby targets offer a wide range in parameter space to %measure feedback efficiency 
quantify feedback efficiency via, for example, the outflow mass-loading factor and the stellar winds and supernovae thermalization efficiency constrained from the hot gas properties.  These properties will be determined as a function of cluster mass, age, metallicity, galactocentric radius, and star cluster concentration, which are necessary inputs to test star and galaxy formation models. 

X-rays are a longstanding missing piece in the puzzle of stellar feedback and its role in galactic ecosystems. By measuring the internal energy of a vast range of star clusters in nearby galaxies for the first time as part of the core science program, AXIS will %offer a much more complete view of star clusters in nearby galaxies,
finally provide the long-sought quantification  of how hot gas from stellar winds and supernovae, X-ray binaries, and cosmic rays regulate star formation and shape galaxies.

\section{Black hole feedback: quasar mode}
\subsection{Quasar-mode feedback in nearby galaxies}
\label{sec:nearbyquasars}
\vspace{-0.5\baselineskip}
\noindent (Contributed by A. Trindade-Falc\~{a}o, H.R. Russell, F. Tombesi and W.P. Maksym)
\vspace{0.5\baselineskip}

\begin{figure}
    \centering
    \includegraphics[width=0.9\columnwidth]{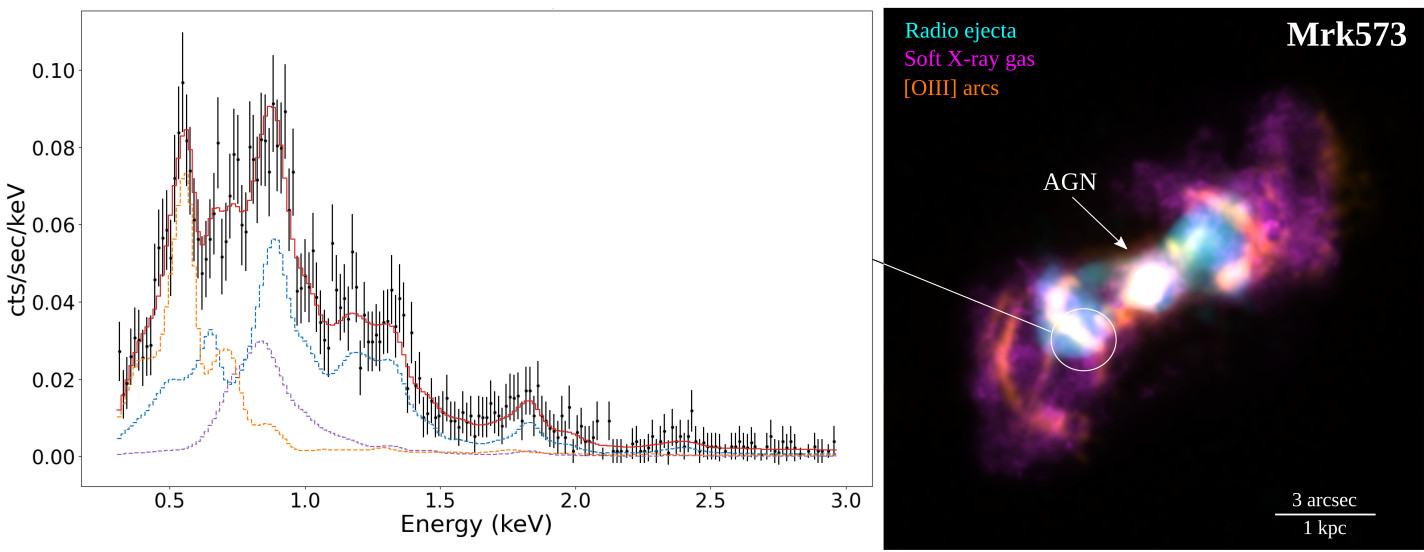}
    \caption[Composite image and AXIS simulation of Mrk573]{Composite image of the Seyfert galaxy Mrk573 comprising a PSF-deconvolved \textit{Chandra} image, radio VLA $6\cm$ emission from the radio jets and HST [O~\textsc{iii}] emission \cite{Paggi12}.  Simulated AXIS spectrum ($150\ks$) for the region shown.  The best-fit model has one thermal component (purple line) and two photoionization components (blue and orange lines).  The total model is shown by the red line.}
    \label{fig:mrk573}
\end{figure}

\noindent Even though the central SMBH has a mass of only $\sim0.1\%$ of its host galaxy, the enormous radiative and mechanical energy released as it grows (and shines as an AGN) can slow or completely suspend the entire galaxy's growth [for a review see \citealp{Fabian12}].  Now included as standard physics in simulations of structure formation, this mechanism may be able to transform gas-rich, star-forming progenitors into `red and dead' ellipticals and imprint the observed strong correlation between black hole mass and galaxy bulge velocity dispersion \cite{Magorrian98,Bower06,Croton06}. The majority of black hole growth occurs during phases of high-accretion radiative/quasar mode where intense radiation and high velocity winds off the accretion disk heat, ionize, and expel gas from the surrounding ISM \cite{Silk98,Fabian99,King03,Murray05}.  However, estimates of the energy and momentum input vary widely, and the necessary coupling to kpc-scale cold gas clouds is still far from being fully understood.

Spatially-resolved spectroscopy on scales of hundreds of parsecs (arcsec scales) is required to resolve the intricate structure of shocks, and ionization cones, to map the resulting transitions in the hot gas properties, and thereby to measure the energy transmitted to the surrounding ISM.  With an order of magnitude increase in soft X-ray effective area, AXIS will dramatically extend our reach beyond the few brightest, most massive systems to cover extensive samples of individual galaxies in the local universe and to develop physical models for the unresolved quasar feedback at the peak epochs of galaxy and black hole growth at much higher cosmological redshifts (z$>$2).

AXIS observations of nearby Seyferts will map the impact zones where winds from the luminous AGN first meet the ISM (Fig. \ref{fig:mrk573}).  Outflow velocities at hundreds of parsec scales are typically $\sim1000\kmps$, for which the characteristic interaction energies are $1\keV$ and require X-ray observations. Soft X-ray bright knots and arcs result from a complex mixture of photoionization and thermal shock emission [e.g., \citealp{Fabbiano18,Maksym19,TrindadeFalcao23}], which AXIS will disentangle with multi-component spectral models on arcsec scales and measurements of multiple high-ionization species to break degeneracies.  With its unprecedented large effective area, AXIS will detect very low surface brightness components to test complex shock heating models, such as "slow cooling" where the gas is strongly heated, cannot cool efficiently, and produces disconnected arcs of X-ray emission as the energy is dissipated far beyond the deceleration point [e.g., \citealp{Fornasini22}].  AXIS could rapidly observe large samples in the GO phase and cover a wide range of AGN luminosities, accretion rates, and galaxy masses. AXIS will then quantify the thermal power injected into the ISM as a fraction of the AGN bolometric luminosity and test different feedback models.  Strong coupling of the radiative energy would imply so-called simple AGN feedback, where a supersonic pressure or momentum-driven outflow off the accretion disk propagates to large scales.  Weaker coupling would instead signify a two-stage mechanism where radiative pressure drives a weak wind in the hot ISM and secondary effects, such as the radiative effects of dust absorption and ionization, act on cold gas at large scales \cite{King15}.

\subsection{Quasar feedback at high cosmological redshifts}
\vspace{-0.5\baselineskip}
\noindent (Contributed by G. Chartas)
\vspace{0.5\baselineskip}

\begin{figure*}
\begin{minipage}{\textwidth}
\centering
\includegraphics[scale=0.75]{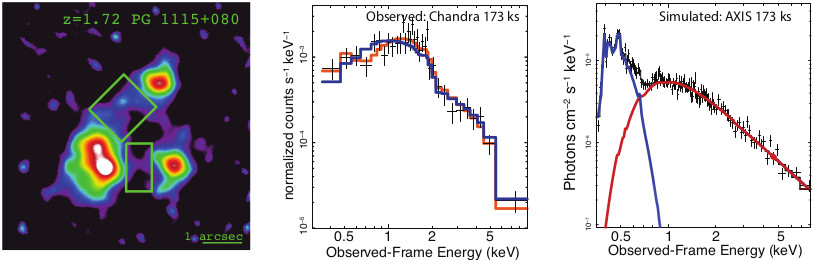}
\caption[A partial X-ray Einstein ring and simulated AXIS spectrum for PG1115+080]{Left: A deconvolved \textit{Chandra} image obtained by stacking all available observations of PG1115+080 (total exposure time of $173\ks$). A partial X-ray Einstein ring is resolved between lensed images. Center: The stacked $173\ks$ \textit{Chandra} spectrum extracted from the regions (left) covering the partial X-ray Einstein ring.  A model that includes a plane-parallel shock plasma component plus a power-law component is shown in red. The spectrum can be equivalently fit by only a power-law model (blue). Right: A simulated $173\ks$ unfolded AXIS spectrum of the partial X-ray Einstein ring clearly resolves the shocked plasma component (blue) from the power-law component (red).}
\label{fig:einstein_rings}
\end{minipage}
\end{figure*}

\noindent The observed evolution of the stellar mass of galaxies [e.g., \citealp{Madau2014}]  indicates that approximately 75\% of the stellar mass had already formed before $z \sim 1.3$. It is therefore important to determine the energetics and morphology of multi-phase outflows/winds at high redshift, near the peak of AGN and star formation activity. One prediction of numerical models of ultrafast outflows (UFOs) is that they interact with the ISM, producing shocks and extended X-ray emission at the shock front [e.g., \citealp{Faucher2012,Zubovas2012}]. The extent of the shocked ISM gas is predicted to be of the order of $\sim 0.1-1\kpc$. At $z = 2$ , this corresponds to a range of angular sizes of $0.01 - 0.1\asec$ that cannot be resolved by any current X-ray telescope.  A direct detection of the interaction of a small-scale UFO originating at $10$-$100$ $r_{\rm g}$, where $r_{\rm g}$ = G$M_{\rm BH}$/c$^{2}$, with kpc-scale gas would provide confirmation that UFOs transfer a significant amount of their energy to the host galaxy and are thus an important component of galaxy feedback.

When gravitationally lensed, a compact quasar source forms multiple images whilst any extended emission around the quasar forms a ring (aka Einstein ring) that connects the individual images. A technique that has been successfully employed to resolve extended optical and IR emission around quasars involves lens modeling the Einstein rings of gravitationally lensed quasars. Finding lensed extended emission is easier than finding unlensed extended emission because the lens magnification enormously reduces the contrast between the extended emission and the central quasar \cite{Kochanek2001}. The detections of X-ray Einstein rings associated with UFOs in distant lensed quasars are rare due to their X-ray weakness and the limited number of known lensed quasars with relativistic outflows \cite{Chartas2021}.
The detection of both the UFO from the spectra of the lensed images and the extended emission from the shock can only be accomplished with an X-ray telescope with a PSF of the order of 1 arcsec (required to resolve the lensed images and Einstein ring) and an effective area of about $>$ 10 times \textit{Chandra} (required to obtain high signal-to-noise UFO and shocked ISM spectra). AXIS is the only proposed X-ray probe mission that meets these requirements.
 
A hint of the detection of both a UFO and an Einstein ring is provided in \textit{Chandra} observations of the lensed quasar PG 1115+080 \cite{Chartas23}.  PG 1115+080 is a quadruply lensed quasar containing a powerful UFO ($v\sim0.4$c). Our preliminary spatial analysis of  the available  \textit{Chandra} observations of PG1115+080 shows hints of a partial X-ray Einstein ring. In Fig. \ref{fig:einstein_rings}, we show the deconvolved \textit{Chandra} image obtained by stacking all available \textit{Chandra} observations (total exposure time of $173\ks$). The extended emission forming a partial Einstein ring is visible in both the stacked observation and in a single $30\ks$ observation of PG115+080. The \textit{Chandra} spectrum of the partial Einstein ring is also shown in Fig. \ref{fig:einstein_rings}. An acceptable spectral model includes a plane-parallel shock plasma model and an absorbed power-law model. Due to the low signal-to-noise ratio, however, a model that only includes an absorbed power law also provides a satisfactory fit to the data.  
We simulated an AXIS spectrum with a matching exposure time ($173\ks$) using the best-fit model to the \textit{Chandra} spectrum of the partial Einstein ring (Fig. \ref{fig:einstein_rings}). The simulated AXIS observation of PG1115+080 shows that we will unambiguously detect the presence of shocked emission from the interaction of the UFO with the ISM.

\begin{SCfigure}
\includegraphics[scale=0.35]{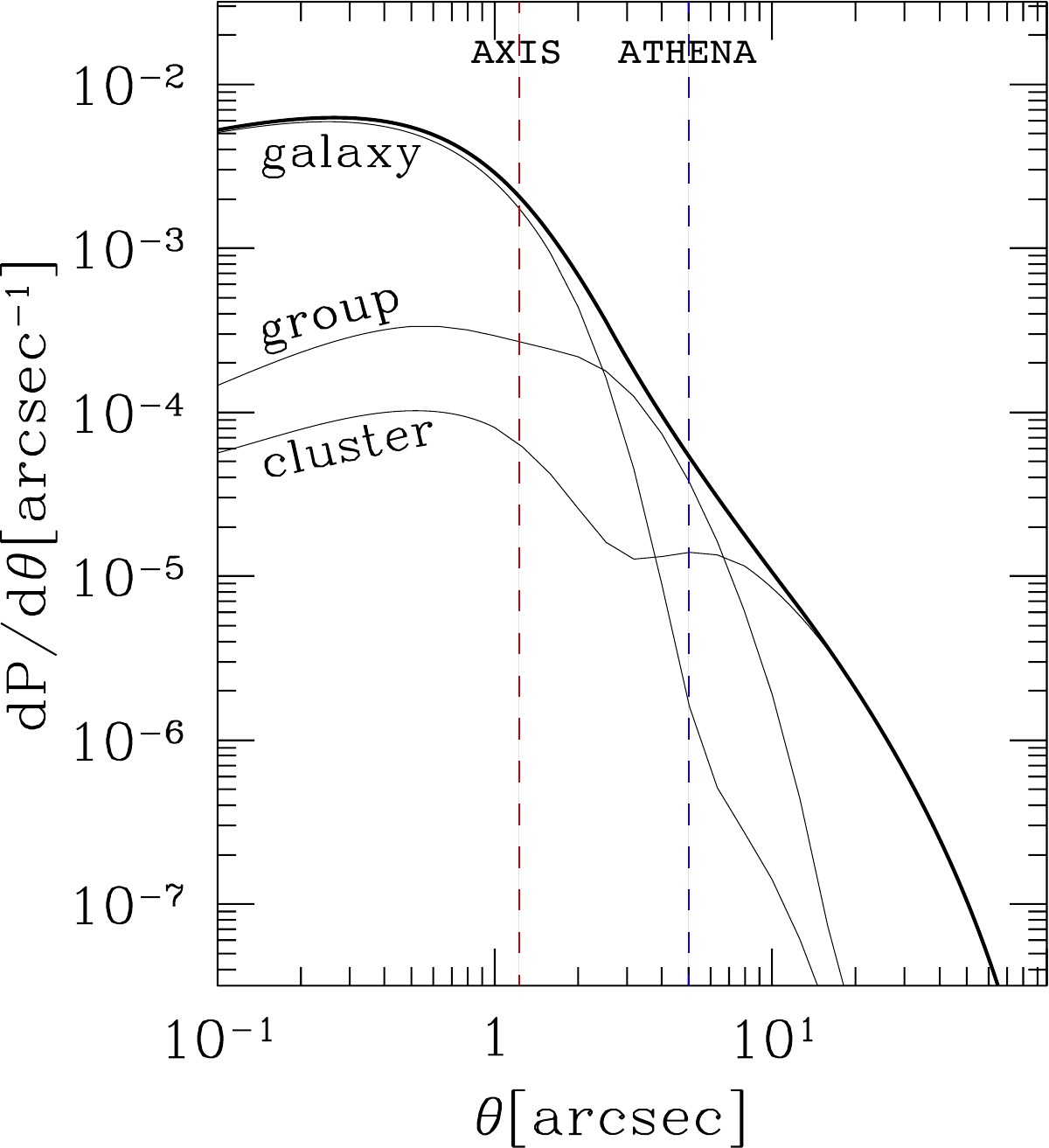}
\caption[The distribution of lens image separations compared to the AXIS half-power diameter]{\protect\rule{0ex}{6ex}The distribution of lens image separations for three different lens types: galaxy, group, and cluster-scales, predicted by a halo model. The total distribution is shown by the thick line. The vertical lines represent the current best estimate half-power diameters of AXIS and Athena.}
\label{fig:image_separations}
\end{SCfigure}

There is an important synergy between AXIS and other multi-wavelength observatories that will improve our understanding  of galaxy feedback through multi-phase winds. Spectra and images from mm observatories (e.g., ALMA) will be used to determine the properties of the macro-scale winds. Numerical models of UFOs [e.g., \citealp{Faucher2012,Zubovas2012}] predict that they drive powerful winds that heat and expel cold gas from the galaxy on larger scales.  At present, there are few examples where the momentum flux of the AGN wind can be linked to a large-scale molecular outflow, but a combination of AXIS and ALMA can increase the sample and allow us to test models of this process. The Vera C. Rubin Observatory is predicted to discover $>10,000$ gravitationally lensed quasars, most of which will be resolved with AXIS [\citealp{Oguri2006}, see Fig. \ref{fig:image_separations}]. The planned observing window of AXIS ($\sim2032-2037$) will overlap with the multi-band photometric optical surveys of Rubin ($\sim2024-2034$), GAIA and EUCLID.  The eROSITA all-sky survey will provide the X-ray fluxes of the lensed quasars discovered by Rubin and other surveys (J-PAS, Pan-STARRS1, DESI legacy, and Dark Energy Survey) but cannot resolve the lensed images or extended emission.  AXIS will target lensed systems that are X-ray bright enough and detect UFOs and Einstein rings.

AXIS observations of a sample of high-$z$ lensed quasars will allow us to study both the central drivers of galaxy growth at the smallest scales ($1-100 r_{\rm g}$)  and the interaction of the winds with the ISM at mesoscale ($\sim\pc-\kpc$).

\section{Black hole feedback: radio mode}

\subsection{Radio-mode feedback in galaxies}
\vspace{-0.5\baselineskip}
\noindent (Contributed by L. Lanz and H.R. Russell)
\vspace{0.5\baselineskip}
\label{sec:radiomodegals}

\begin{figure}
    \centering
    \raisebox{0.5cm}{\includegraphics[width=0.53\columnwidth]{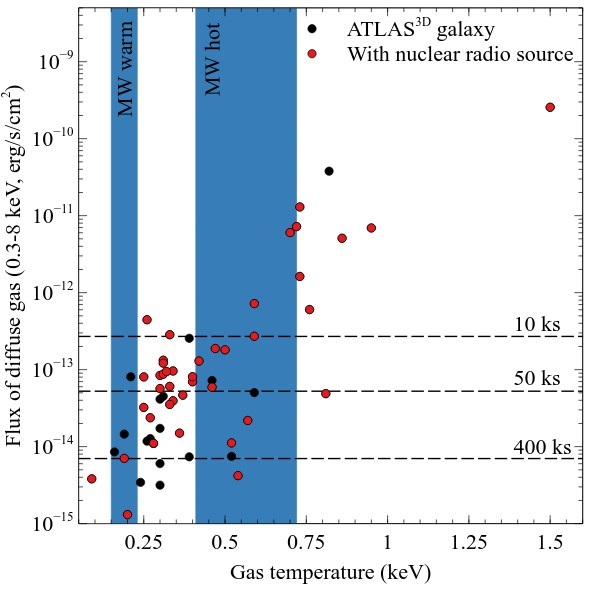}}
    \includegraphics[width=0.43\columnwidth]{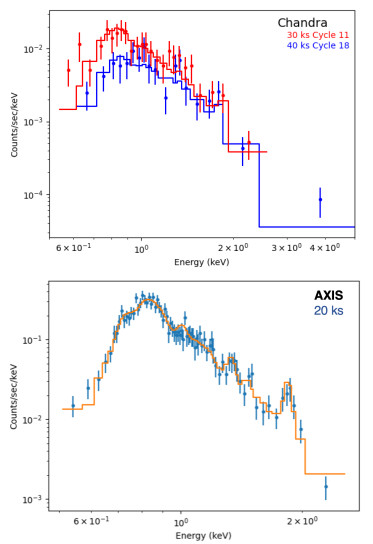}
    \caption[The AXIS view of Atlas$^{\mathrm{3D}}$ galaxies and simulated spectra for NGC\,1266]{Left: X-ray flux vs. gas temperature for galaxies in the ATLAS$^{\mathrm{3D}}$ sample \cite{Cappellari11} with {\em Chandra} observations \cite{Kim15}.  AXIS exposure times to obtain $>10,000$ counts from the diffuse gas are shown for representative targets.  Note that, given the low temperatures of these hot atmospheres, $10\ks$ with AXIS is equivalent to $250\ks$ with \textit{Chandra} at launch and $2.5\Ms$ currently.  Uncertainties on \textit{Chandra's} global temperature measurements are not shown for clarity, but are typically $15-80\%$.  Galaxies with and without detected nuclear radio emission are shown by the red and black points, respectively \cite{Nyland16}.  The temperature components of the Milky Way's circumgalactic medium are shown for reference [e.g. \citealp{Das19}].  Right: {\em Chandra} (upper) and simulated AXIS spectra (lower) of NGC\,1266, a post-starburst lenticular with a 6$''$ multiphase nuclear outflow, showing the degradation of {\em Chandra}'s soft response and the benefit of AXIS' high effective area. More than twice as many photons were collected in less than 1/7th of the exposure time. }
    \label{fig:ngc1266}
\end{figure}
\begin{figure}
    \centering
    \includegraphics[width=0.95\columnwidth]{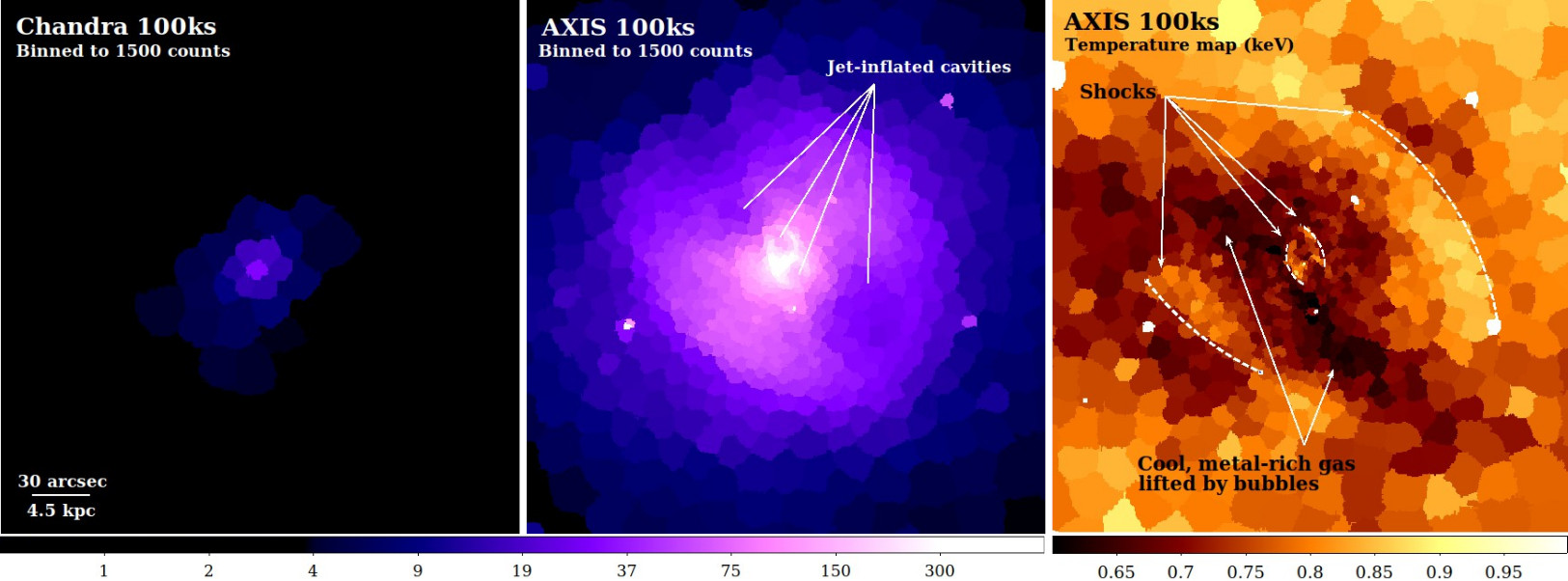}
    \caption[A \textit{Chandra} image and an AXIS simulation of AGN feedback in NGC\,5813]{Left and center: \textit{Chandra} (at launch) and AXIS simulated images of AGN feedback in NGC\,5813 ($100\ks$) with spatial binning for 1500 counts per region.  Color bars are matched and units are $\ctspasecsq$.  The radio jet has carved out large cavities in the galaxy's hot atmosphere, which are visible in the AXIS simulation as depressions in X-ray surface brightness.  With 1500 counts per region, AXIS will map the gas properties on these spatial scales to an accuracy of e.g. a few percent in temperature.  The spatial bins were generated with the contour binning algorithm and group pixels with similar surface brightness \cite{Sanders06}. All panels cover the same field of view.  Right: Simulated temperature map in keV for a $100\ks$ AXIS observation, which probes the thermodynamic properties on the scales of the bubble rims, shocks, and cool gas plumes \cite{Randall11,Randall15}.}
    \label{fig:ngc5813}
\end{figure}

\noindent Powerful radio jets are a means by which AGN exert influence on their host galaxies, but how they are able to provide steady, distributed, and self-regulated heating to galaxy and cluster environments is still not well understood [see, e.g., \citealp{Fabian12,McNamara12} for reviews]. Mechanical- or radio-mode feedback typically occurs when accretion is happening at low rates ($\lesssim1\%$ of the Eddington rate), enabling this mode to be active for longer periods than intense quasar mode feedback (see section \ref{sec:nearbyquasars}).  Radio-mode AGN have primarily been studied in massive systems, where they must efficiently couple to their surrounding hot X-ray atmospheres to suppress cooling and maintain low levels of star formation at late times. The physical processes by which the energy and momentum are transferred to cooling and star-forming gas, distributed on galaxy scales, and how these mechanisms scale with host mass remain unknown.

Spatially-resolved spectroscopy with high angular resolution is critical for resolving and characterizing extended diffuse emission and differentiating between feedback from nuclear star clusters and star formation and feedback from the AGN including via small-scale jets. As jets encounter the ISM, they can shock some of this gas to a range of temperatures, producing emission in the mid-infrared [e.g., \citealp{ogle14}] and X-ray bands [e.g., \citealp{lanz15}]. AXIS's high throughput coupled to arcsecond spatial resolution will be key in resolving the structure and properties of the hot gas for comparison with warm gas mid-IR and near-IR tracers mapped by JWST and Roman, enabling us to trace the processes by which energy is deposited into the ISM from radio jets.  Fig. \ref{fig:ngc1266} shows a comparison of {\em Chandra} observations from Cycle 11 and Cycle 18 with an AXIS spectrum of the same outflow region in NGC\,1266. The typical temperature of galactic hot atmospheres lies below $1\keV$ and the dramatic loss of \textit{Chandra} ACIS soft response (currently only $8\cmsq$ at $0.5\keV$) has long placed these targets out of reach [e.g., \citealt{Werner19}].  With an effective area more than 400 times \textit{Chandra} ACIS at $0.5\keV$, AXIS will allow detailed studies of the mechanical processes by which AGN impart energy to their host galaxies and crucially widen our view from individual investigations of only the brightest sources to statistically important samples.

On larger scales of a few to tens of kpc (arcsec to tens of arcsec), jets inflate huge bubbles of radio-emitting plasma and couple very efficiently to the volume-filling hot CGM [e.g., \citealp{McNamara00,Fabian00,Churazov01,McNamara07}].  Our current understanding is almost entirely shaped by deep \textit{Chandra} observations of the nearest, richest cluster atmospheres [e.g., Perseus, \citealp{Fabian06}; M87, \citealp{Forman07}; Centaurus, \citealp{Sanders16}; Abell\,2052, \citealp{Blanton11}].  Although remarkable, these systems are clearly unrepresentative of the wider galaxy population driving the slowdown in star formation at low redshift.  \textit{Chandra} observations of the brightest ellipticals indicate analogous jet interactions with the galaxy's hot atmosphere \cite{Allen06,Nulsen09,Panagoulia14} but have been severely limited by the loss of ACIS's effective area at low energies.  AXIS's expansive soft response will enable Perseus-like studies for numerous systems in the nearby universe to determine if feedback stabilizes cooling in all nearby massive galaxies, whether cluster processes scale down, and what the dominant mechanism is distributing and dissipating energy throughout the galaxy's hot atmosphere.  Fig. \ref{fig:ngc1266} (left) shows all galaxies in the ATLAS$^{\mathrm{3D}}$ sample with {\em Chandra} observations \cite{Kim15} measuring the global properties of the hot atmosphere.  With at least 10,000 counts per target in a large sample with AXIS, we will identify cavities, measure the gas properties in radial profiles and calculate the energy input by the AGN.  AXIS observations of massive ellipticals (Fig. \ref{fig:ngc5813}) will uncover additional pairs of relic bubbles from previous AGN outbursts to reveal the history of AGN activity, whether bubbles break up, merge or pile up at large radius, trace metal-rich, low entropy gas flows, and determine the implications for the distribution of the energy input by jets.  In this way, for the first time, AXIS will reveal how feedback operates across the whole gamut of hot atmospheres from L* galaxies at $10^{12}\Msun$ to the richest clusters above $10^{15}\Msun$. 

\subsection{Feedback in galaxy clusters}
\vspace{-0.5\baselineskip}
\noindent (Contributed by Y. Qiu)
\vspace{0.5\baselineskip}
\label{sec:clusterfeedback}

\begin{figure*}
\begin{minipage}{\textwidth}
\centering
\includegraphics[width=0.95\columnwidth]{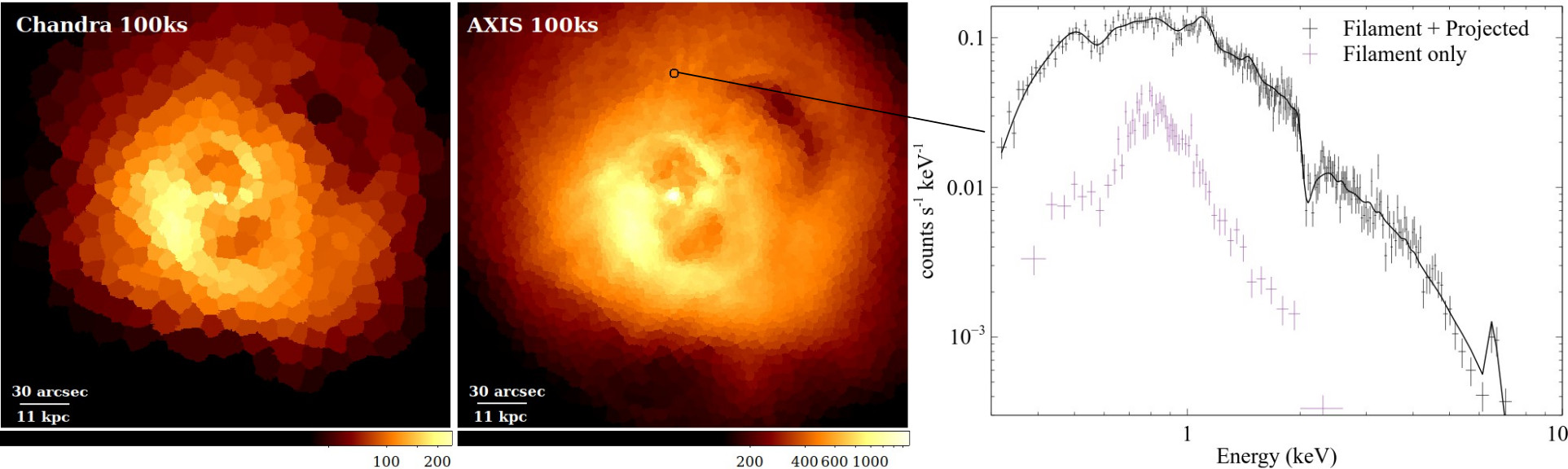}
\caption[A \textit{Chandra} image and AXIS simulations of AGN feedback in the Perseus cluster]{Left: \textit{Chandra} image of the Perseus cluster ($100\ks$ at launch) with spatial binning for 7500 counts per region (color bar units counts arcsec$^{-2}$).  Center: AXIS simulated image of the Perseus cluster ($100\ks$) also binned for 7500 counts per region.  Right: AXIS simulated spectra for the region shown with a radius of $2.5\asec$ (factor of 10 reduction in spatial bin size compared to \textit{Chandra}).  The total spectrum and model are shown in black, and the intrinsic emission from a region of the soft X-ray filament [shown center, projected emission subtracted with a neighboring off-filament region, \citealp{Fabian11filament}] is shown in purple.}
\label{fig:perseusfils}
\end{minipage}
\end{figure*}

\noindent AGN feedback reaches far beyond the SMBH's sphere of influence to interact with the multiphase medium in and around galaxies. In galaxy clusters, where there is a concentration of hot gas pulled in by the dark matter halo, AGN feedback acts against gravity to shape the structure of the X-ray-emitting intracluster medium (ICM). Direct imaging of the ICM with X-ray facilities has proven fruitful in detecting imprints of AGN activity. For example, Chandra has detected the innermost cavities inflated by radio jets in numerous rich clusters and shown that the energy required to inflate them is sufficient to suppress cooling of the cluster atmospheres \cite{Birzan2004, McNamara07}. However, the physical mechanisms by which the jets heat the ICM or the residual gas cooling fuels the AGN remain unknown. Very deep \textit{Chandra} observations of a few of the nearest clusters, such as Perseus, reveal highly structured networks of soft X-ray filaments, weak shocks, sound waves, turbulence and numerous pairs of outer bubbles [e.g., \citealp{Fabian2003}]. AXIS's angular resolution, soft response, and wide field of view are therefore essential to understand how the distribution, multi-temperature thermal state, and metallicity evolution of the plasma are influenced by AGN feedback in a wide range of clusters. 

The cooling of the ICM in cool-core clusters is inextricably linked with AGN feedback [e.g., \citealp{Li2015, Qiu2018, Qiu2019}]. How the plasma cools and forms (sometimes star-forming) cold filaments is still a highly debated topic, including theoretical models such as plasma thermal instability, uplift of low-entropy gas by AGN-driven bubbles, as well as radiatively cooling AGN-driven outflows \cite{Pizzolato2005, Gaspari2012, Li2014, McNamara2016, Voit2017, Qiu2020, Qiu2021}. AXIS's soft response and spatial resolution are perfect for tracing the lowest energy plasma out of which these filaments form (Fig. \ref{fig:perseusfils}), potentially unveiling the fueling of the central AGN and the growth of the stellar component. When combined with optical and radio observations (such as SITELLE, MUSE, and ALMA) that capture the velocity distribution and turbulent structures of the cold gas \cite{Li2020, Hu2022, Zhang2022}, AXIS will enable a comprehensive account for the continuous thermodynamical evolution of the plasma, and further reveal the role AGNs play in tightly controlling the growth of these elliptical galaxies.

\begin{figure*}
\begin{minipage}{\textwidth}
\centering
\includegraphics[width=0.32\columnwidth]{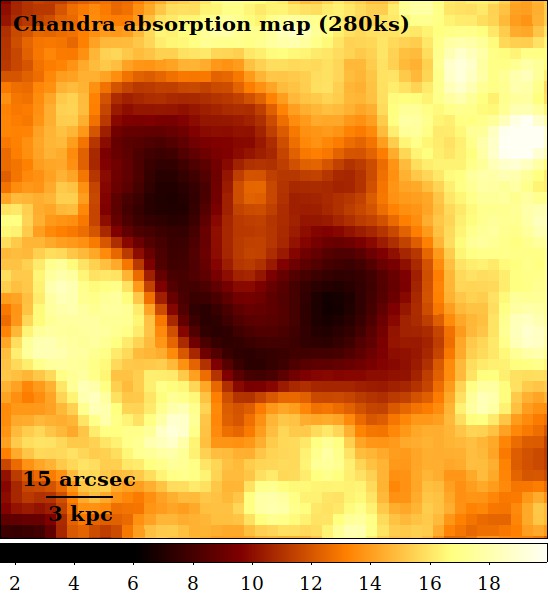}
\includegraphics[width=0.32\columnwidth]{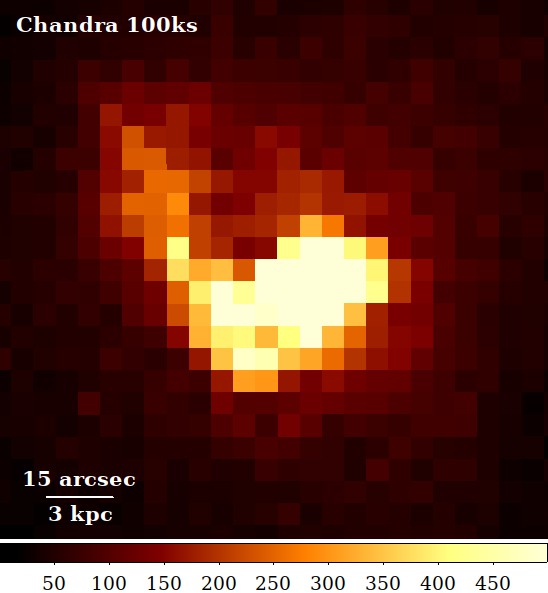}
\includegraphics[width=0.32\columnwidth]{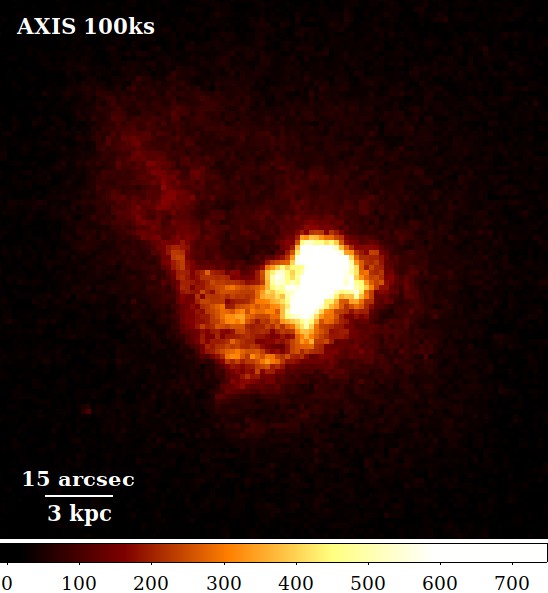}
\caption[\textit{Chandra} images and AXIS simulations of strong absorption at the center of the Centaurus cluster]{Left: Ratio of soft ($0.5-0.7\keV$) and mid-band ($0.8-1.1\keV$) \textit{Chandra} images ($280\ks$ taken $2000-2004$) showing regions of strong absorption in the center of the Centaurus cluster.  Center: $100\ks$ \textit{Chandra} soft band counts image binned for $>100$ counts per pixel in regions of interest.  Right: Same as center but for $100\ks$ with AXIS.  Substantial expansion of the soft response with AXIS ensures over an order of magnitude improvement in resolution.}
\label{fig:hcf}
\end{minipage}
\end{figure*}
\begin{figure}
 \begin{minipage}{\textwidth}
\centering
\raisebox{0.75cm}{\includegraphics[scale=0.47]{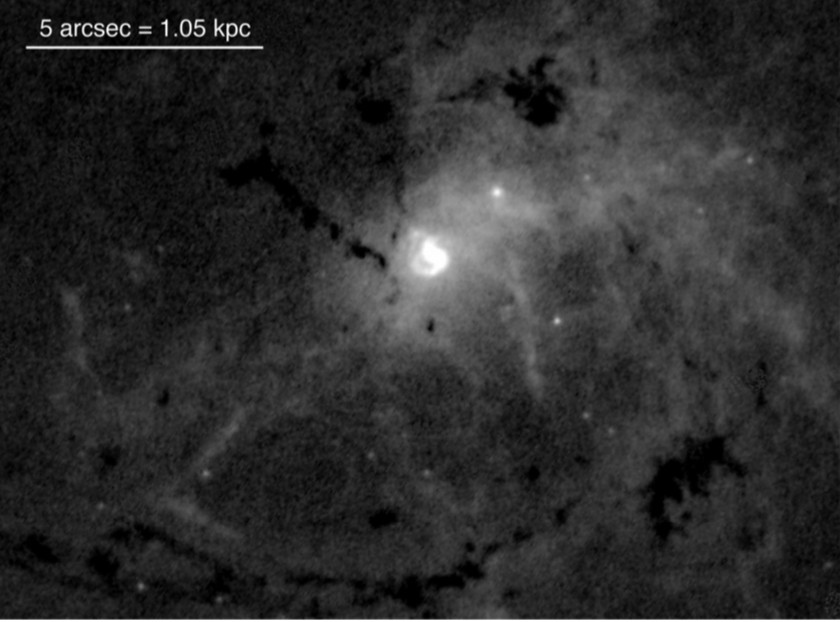}}
\includegraphics[scale=0.31]{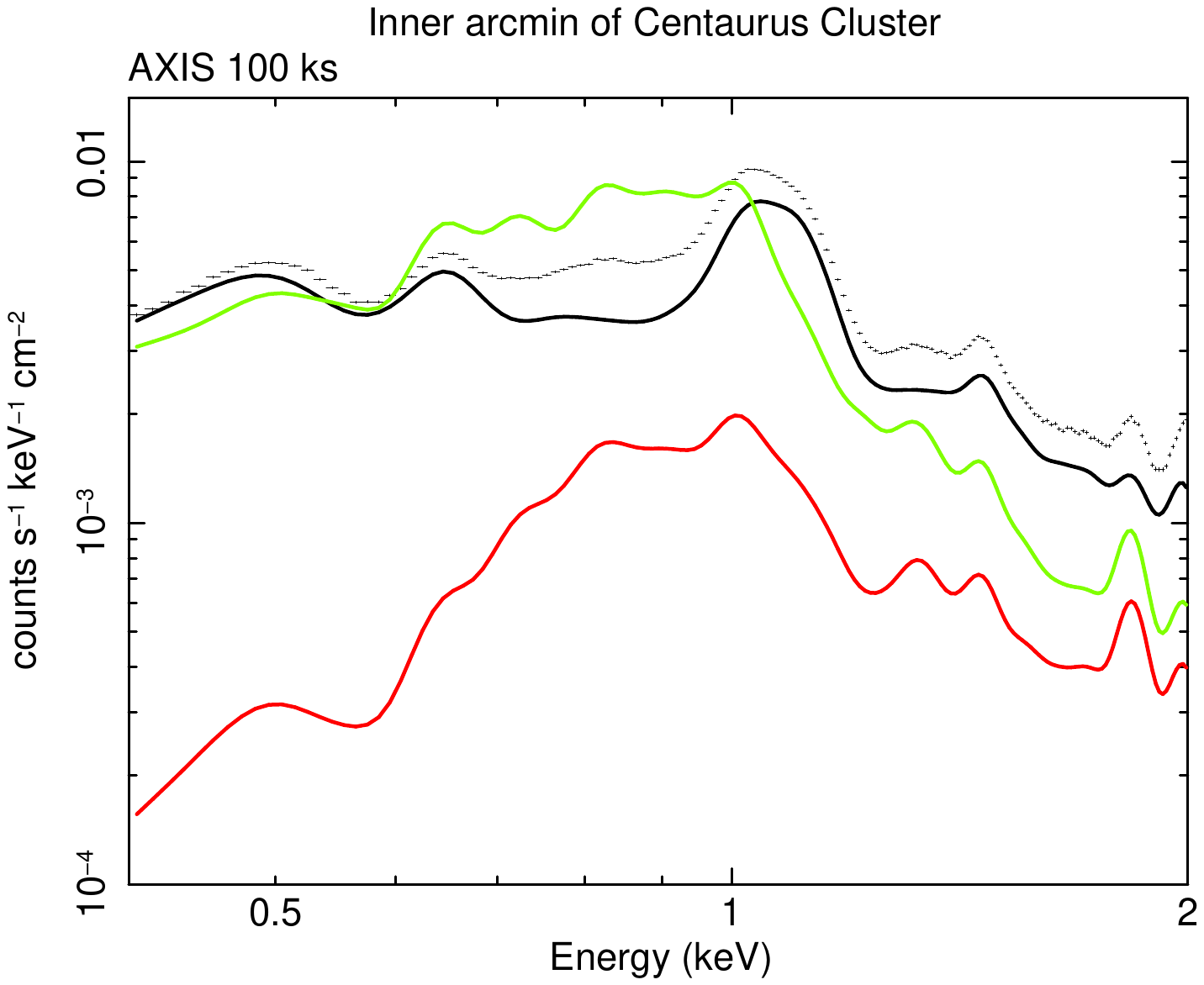}
\caption[An HST image of dusty filaments in the Centaurus cluster and an AXIS simulation of a hidden cooling flow spectrum.]{Left: HST H$\alpha$ image of the core of the nearby Centaurus Cluster around its Brightest Cluster Galaxy NGC\,4696 \cite{Fabian16}. Note the dusty filaments and emission clouds, all of which will absorb soft X-rays originating from this region. Right: Simulated $100\ks$ AXIS spectrum of the inner arcmin diameter of NGC\,4696. The green line shows a $15\Msunpyr$ cooling flow, the red line similar but with a total intrinsic absorption of $1.6\times10^{22}\pcmsq$ included to match the XMM RGS result \cite{Fabian22}. The AXIS spectrum will vary between the green to below the red line depending on the absorption along a specific sightline, as in left.} %Below: HST color composite images of BCGs in RXJ1532.9 (left) and MACS J1931 (right), at $z=0.363$ and 0.352, respectively (ref). Both have massive hidden cooling flows. RXJ1532 has a lower Galactic column density and similar observed central X-ray surface brightness as  Centaurus so will yield comparable AXIS spectra over the  inner 10 arcsec.}
\label{fig:N4696}
\end{minipage}   
\end{figure}

In the absence of AGN feedback, the properties of the ICM, such as temperature and density distribution, are determined primarily by the underlying gravity \cite{Voit2005}. Detecting and characterizing property modifications associated with activities of the AGN are therefore crucial in understanding how feedback energy is delivered to the surrounding medium. Measurements of shocks, cavities, and turbulence [e.g., \citealp{Birzan2004, Forman2007, Zhuravleva2014}] in the plasma all require a large collecting area with arcsecond resolution. AXIS's advancement over existing X-ray imaging telescopes enables the detection of these features beyond the central tens of kpc. The combination of spatial resolution (providing detailed maps of the density and temperature structure) with velocity structure from future X-ray calorimeters (such as XRISM, HUBS, and Athena) is key to understanding how energy is transmitted to the ICM and distributed throughout the cluster core.

\subsection{Hidden cooling flows in clusters}
\vspace{-0.5\baselineskip}
\noindent (Contributed by A.C. Fabian)
\vspace{0.5\baselineskip}

\noindent The radiative cooling time at the center of the hot atmospheres of massive galaxies, groups and clusters can drop below $10\Myr$.  The observed mass cooling rate of this gas is very low and drops to near zero as the gas temperature falls below $0.4\keV$ \cite{Peterson01,Sanders08}.  Either AGN feedback is very tightly balanced, as discussed in sections \ref{sec:radiomodegals} and \ref{sec:clusterfeedback}, or the soft X-ray emission from cooling is somehow hidden from view, which is considered here.  The gas in the centers of cool core clusters, groups, and elliptical galaxies is then cooling at significant rates of order 100, 10 and $1\Msunpyr$, respectively. As the gas cools below $1\keV$ in the central kpcs it becomes entwined with the gas that has already cooled. It is hidden from direct detection by photoelectric absorption in the cold gas, as revealed by a recent analysis of \textit{XMM-Newton} RGS spectra \cite{Fabian22,Fabian23}.

The fate of the rapidly accumulating cold gas is unknown, but the observed metallicity peaks at about $10\kpc$ radius in nearby clusters indicate slow outflows of enriched gas from their centers. Some gas may, however, become ultra-cool at close to $3\K$, fragment and form low mass stars \cite{Fabian22}.  Some low-mass stars may be swallowed whole by the central black hole while emitting little radiation \cite{Fabian23b}.

The high angular resolution and soft response of AXIS will enable detailed spatial (Fig. \ref{fig:hcf}) and spectral (Fig. \ref{fig:N4696}) mapping of these important regions, uncovering the workings of the innermost parts of the most massive galaxies in the Universe. 
  
\subsection{Inverse Compton Ghosts in the deep X-ray Sky}
\vspace{-0.5\baselineskip}
\noindent (Contributed by A.C. Fabian)
\vspace{0.5\baselineskip}

\begin{figure}
\begin{minipage}{\textwidth}
\centering
\raisebox{0.75cm}{\includegraphics[scale=0.45]{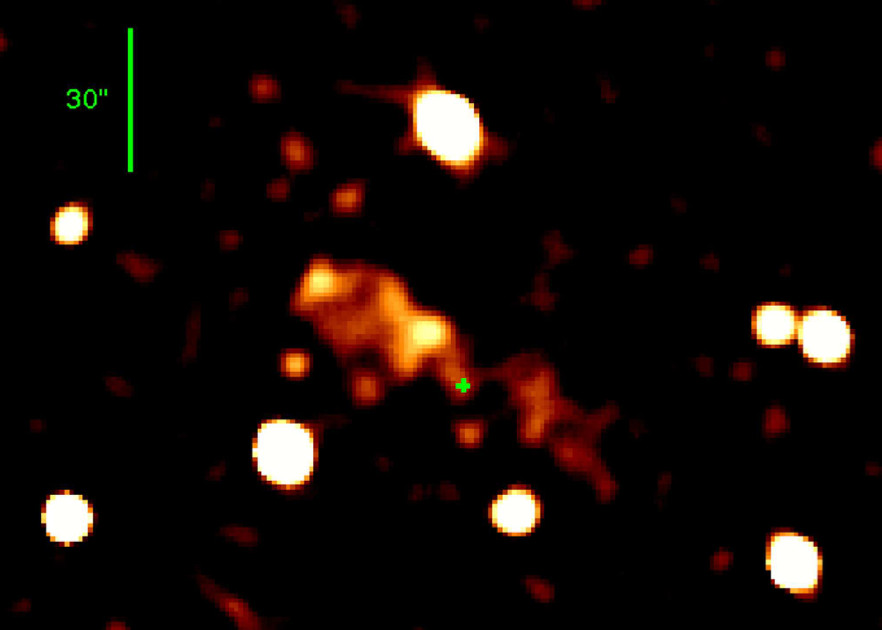}}
\includegraphics[width=0.48\textwidth]{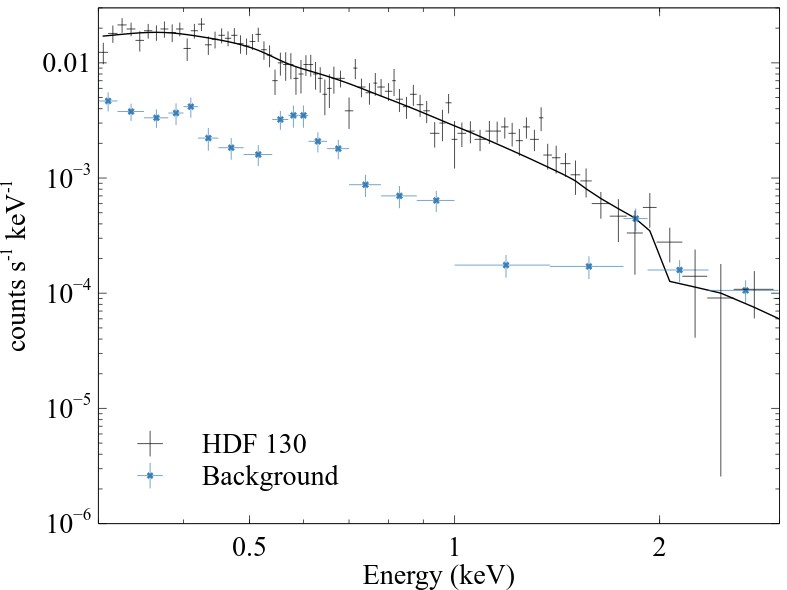}
\caption[A \textit{Chandra} image and a simulated AXIS spectrum of the inverse Compton ghost HDF 130]{Left: The extended X-ray emission around HDF 130 at $z=1.99$ \cite{Fabian09}. It is an inverse Compton ghost of a giant radio source in the \textit{Chandra} Deep Field North ($2\Ms$). The \textit{Chandra} spectrum for HDF 130 can be equivalently fit by power-law and thermal models.  Right: Simulated AXIS spectrum of HDF 130 in a $300\ks$ pointing from the AXIS Wide Survey.  The AXIS simulated background (soft and non-X-ray background) and the best-fit power-law model with $\Gamma=2.65\pm0.05$ and $\chi^{2}=80$ for 89 degrees of freedom (compared to $\chi^2=111$ for 89 degrees of freedom for a thermal model) are also shown.}
\label{fig:hdf130}
\end{minipage}   
\end{figure}

\noindent Powerful jetted AGN feedback from isolated galaxies leads to giant double-lobed sources in the radio band. The energetic electrons in the lobes lose energy to inverse Compton scattering of the microwave background, the energy density $\epsilon_{\rm cmb}$ of which grows with redshift $z$ as $(1+z)^4$. The lifetime of an electron of Lorentz factor $\gamma$ scales as  $1/(\gamma \epsilon_{\rm cmb})$. Consequently, when the jets switch off, the radio emission, requiring $\gamma\sim 10^4-10^5$, dies away faster than the inverse Compton X-ray emission, which requires $\gamma\sim 10^3$. This results in  a double-lobed inverse Compton ghost appearing in the soft X-ray band (its spectrum is likely steep). A candidate ghost was found in the $2\Ms$ \textit{Chandra} Deep Field North, centered on the massive galaxy HDF 130 at $z=1.99$ [Fig. \ref{fig:hdf130}, see \citealp{Fabian09}]. The lack of any such ghosts in the deeper \textit{Chandra} Deep Field South is likely due to the lack of sensitivity to soft X-rays that had developed due to the buildup of obscuring matter in the optical blocking filter on \textit{Chandra's} ACIS detector. Such sources may have a higher space density than clusters at $z>2$ and $L_{\rm X}>10^{44}\ergps$.

Several other examples of jets with X-ray but not radio emission have been reported from \textit{Chandra} observations \cite{Mocz11}. AXIS should detect many more with its capability to map faint X-ray emission. The distribution with redshift and luminosity will provide a valuable guide to the importance and evolution of jetted feedback in isolated galaxies \cite{Mocz11b,Mocz13}.  

\section{The integrated history of feedback with high-z clusters}

\subsection{Radio-mode feedback in high-z clusters}
\vspace{-0.5\baselineskip}
\noindent (Contributed by S.W.\ Allen, A.B.\ Mantz and M.\ McDonald)
\vspace{0.5\baselineskip}
\label{sec:highzcavities}

\begin{SCfigure}
\includegraphics[scale=0.9]{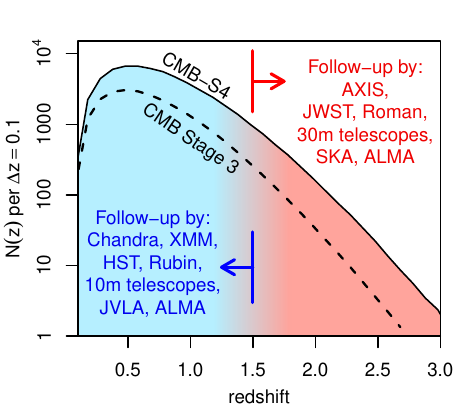}
\caption[SZ cluster detections as a function of redshift]{\protect\rule{0ex}{6ex} The number of SZ cluster detections expected as a function of redshift from Stage~3 SZ surveys and the proposed CMB-S4 project [adapted from \citealp{Mantz19}]. Blue to red shading shows the transition to the $z>1.5$ regime, for which high spatial resolution and throughput are key requirements for extracting information about halo centers, relative masses, dynamical states, internal structure, and galaxy/AGN populations.}
\label{fig:SZdetections}
\end{SCfigure}

\begin{figure*}
\begin{minipage}{\textwidth}
\centering
\includegraphics[width=0.95\textwidth]{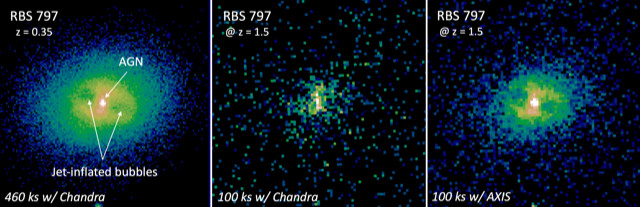}
\caption[AXIS observations of X-ray cavities at high redshift]{Left: \textit{Chandra} observation of the X-ray cavities in RBS\,797.  Two cavities are visible $\sim30\kpc$ to the E and W of the AGN and are each $\sim30\kpc$ across [e.g., \citealp{Ubertosi21}].  Center and right: \textit{Chandra} and AXIS simulations of RBS\,797 scaled for a mass of $M_{500}=5\times10^{14}\Msun$ at $z=1.5$ (applying cosmological dimming, reducing angular size, etc.).  AXIS will extend measurements of AGN feedback in these environments into the high redshift regime for the first time.}
\label{fig:rbs797}
\end{minipage}
\end{figure*}

\noindent Galaxy clusters are the pinnacle of hierarchical structure formation. As the most massive and largest virialized structures in the Universe, clusters provide unique, high density environments within which to study the physics of galaxy evolution \cite{Allen11,Kravtsov2012}. Their deep gravitational potentials preserve the integrated history of generations of star formation and AGN feedback in the thermodynamic properties \cite{McDonald17,Sanders18,Ghirardini21}.   Powerful new surveys at millimeter (e.g., South Pole Telescope, Atacama Cosmology Telescope, Simons Observatory and CMB-S4), optical/near-IR (eg. Rubin, Roman, Euclid) and soft X-ray wavelengths (eROSITA) are now set to expand, by orders of magnitude, the size of cluster catalogs (Fig. \ref{fig:SZdetections}). Critically, the new millimeter and optical/near-IR surveys will also reach much further in redshift, back to the epoch when these massive, virialized structures first formed.  High-spatial-resolution X-ray follow-up observations of high redshift clusters will unfurl the complex interplay between gravity, star formation and AGN feedback over their full history.  However, such measurements lie beyond the reach of existing facilities: detailed studies with \textit{Chandra} and \textit{XMM-Newton} are restricted to $z<1$, with only a handful of low S/N measurements for a few extreme systems that extend to $z\sim1.5$. With its combination of high spatial resolution, expansive soft X-ray response, low instrumental background, and wide field of view, AXIS will transform our understanding of high redshift clusters, allowing us to observe their complete formation history, including the period of ‘cosmic noon’ ($z\sim2-3$), when AGN and stellar activity within them peaked.

AXIS guaranteed time observations will target the most massive, most luminous X-ray clusters discovered at redshifts $1.5 < z < 3$, resolving the AGN within them and measuring the density and temperature of the ICM, to a precision of better than 5\% on spatial scales as small as $20\kpc$. Such measurements, which also provide determinations of the pressure and entropy structure of the ICM, will reveal the evolving and cumulative impact of AGN feedback on its environments. 
%That level of precision will also allow us to robustly assess the dynamical status of z>1 clusters, which are dim and have R$_{200}$ typically less than a few arcmin, currently requiring relaxation to be estimated based on a few morphological quantities from X-ray imaging. The degree of relaxation has a direct impact on recovering their masses through deprojection, minimizing the impact of non-thermal pressure support, global asymmetry and projection effects, that weaken cosmological constraints using clusters \citealp{mantz2014}.
For brighter targets, AXIS will also determine the instantaneous impact of this feedback manifested, for example, by AGN-blown cavities inflated in the surrounding X-ray gas. Such cavities are routinely observed at low-to-intermediate redshifts with \textit{Chandra} and \textit{XMM-Newton}, providing powerful insights into the physics of AGN feedback. High spatial resolution is critical; at redshifts $z>1.5$, these cavities are expected to typically subtend only a few arcsec (Fig. \ref{fig:rbs797}).  AXIS will extend these measurements, for the first time, into the high-redshift regime.

%X-ray measurements of clusters of galaxies played a key role in establishing the modern “concordance” model of cosmology. Clusters are also unique astrophysical laboratories, for which X-ray data provide powerful insights into the physics of galaxy and structure evolution, feedback processes, the history of metal enrichment and the physics of diffuse plasmas. The future of galaxy cluster science is compelling, with cluster catalogs set to expand by orders of magnitude in size and, for the first time, extend out to redshifts z>1.5, when these most massive, virialized structures first formed. Unlocking the full discovery potential of these new cluster catalogs requires a new X-ray observatory, combining high spatial resolution, excellent soft X-ray response and large collecting area. The AXIS mission meets these needs, being capable of resolving the thermodynamic structure within clusters and observing the dominant feedback processes in action during the epoch when star formation and AGN activity peaked. 

%Figures to include: AXIS versions of Fig 1 (combine a and b?) or 2 from Mantz, Allen et al. 2019. https://arxiv.org/pdf/1903.05606.pdf

\subsection{Metallicity evolution}
\vspace{-0.5\baselineskip}
\noindent (Contributed by A.M.\ Flores, A.B.\ Mantz and S.W.\ Allen)
\vspace{0.5\baselineskip}

\begin{SCfigure}
    \includegraphics[scale=0.65]{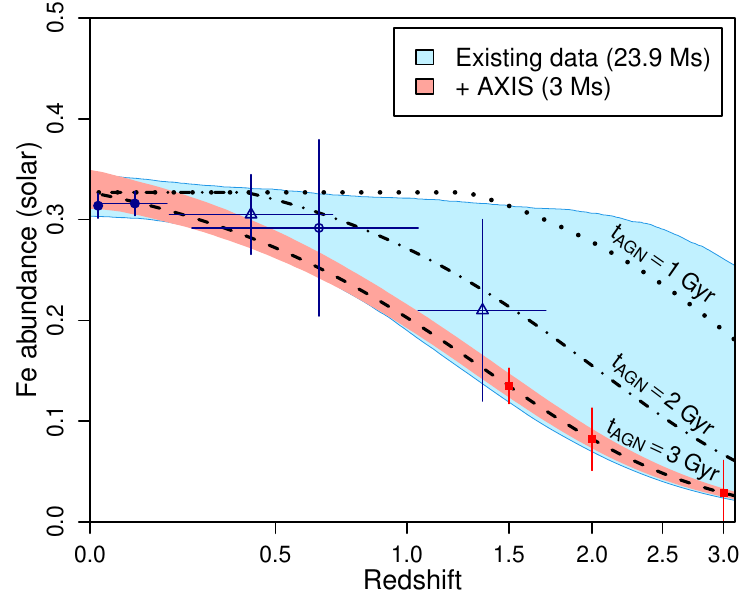}
    \caption[The forecasted improvement in metallicity constraints for cluster outskirts with AXIS.]{\protect 
    The forecasted improvement in metallicity constraints for cluster outskirts with AXIS.
    Current measurements (dark blue points) provide relatively weak constraints on how rapidly enrichment may have advanced in dense cluster environments, remaining consistent with even non-evolving metallicity (light blue band).
    The addition of 3\,Ms of targeted AXIS data at $1.5<z<3$ will provide tight constraints on the rate of enrichment at early times, easily distinguishing between models where the enrichment in dense cluster environments tracks the star formation rate in the field (red points and band), and models where the enrichment of denser environments proceeds faster (dotted and dash-dotted lines).
    $t_\mathrm{AGN}$ labels refer to the age of the Universe when the rate of ICM enrichment is most rapid in each model.
    }
    \label{fig:metal_evo}
\end{SCfigure}

\noindent Measurements of cluster metallicity encode the entire history of star formation and AGN activity within cluster volumes, up to the redshift of observation, providing  unique insights into the enrichment history of the Universe. Metallicity distributions in clusters have been well characterized for nearby systems [\citealp{Werner13,Urban17}] and for statistical ensembles out to z$\sim$1.5 \cite{McDonald16,Flores21}.  These reveal a remarkable universality from system to system and region to region within clusters \cite{Werner13,Urban17}, which requires that AGN feedback played a profound role in redistributing metals throughout the Universe at early times \cite{Biffi1707.05318,Biffi1801.05425,Vogelsberger1707.05318}. However, current X-ray flagship observatories lack the capability to perform such measurements at the redshifts most critical for evolutionary studies, $1.5<z<3.0$, when star formation and AGN activity in the Universe peaked.

In order to probe high redshifts, including the period of cosmic noon ($2<z<3$), where we expect these processes to have been most prolific, we require spatially resolved X-ray measurements of clusters at these redshifts. As discussed in section \ref{sec:highzcavities}, Sunyaev-Zel’dovich (SZ) surveys, from SPT, ACT, the Simons Observatory and CMB-S4, will soon unveil, for the first time, the most massive, virialized systems at these redshifts, which will be prime targets for X-ray followup with AXIS.
Despite being the most massive structures in the Universe, most of the extended emission from clusters at these redshifts is concentrated within regions less than an arcminute in size ($\sim$r$_{500}$). To simultaneously explore the physics of ongoing feedback within cluster cores, and the low surface brightness emission from cluster outskirts, while excising the emission from contaminating point sources, we require an instrument combining high sensitivity, high spatial resolution, and low particle background contamination. Only AXIS provides this combination. 

The proposed $3\Ms$ AXIS observing program for 30 clusters at $1.5<z<3$ will transform our understanding of metal enrichment at high z, providing precise measurements for individual systems and tracing the evolutionary history of cluster enrichment as a whole. In combination with $>20\Ms$ of existing X-ray observations from \textit{Suzaku}, \textit{Chandra}, and \textit{XMM-Newton} at lower redshifts, the new AXIS measurements will provide the first meaningful constraints on the enrichment of the ICM at early times. The data will distinguish, to high precision, models in which enrichment in clusters tracks the star formation rate in the field, from models in which enrichment in the densest environments proceeds at an accelerated pace [as hinted at by recent studies with HST, \citealp{Willis2001.00549}, and JWST, \citealp{Tacchella2302.07234}]. 

Fig.~\ref{fig:metal_evo} shows the existing constraints on the metallicity of cluster outskirts ($0.3 < r/r_{500}<1.0$) from \textit{Suzaku}, \textit{Chandra} and \textit{XMM-Newton}, binned in redshift, and the predicted measurements from 3\,Ms of AXIS data (100\,ks for each of 30 clusters, divided equally among redshifts 1.5, 2.0 and 3.0). The baseline model assumed in the AXIS simulations (dashed line) is one where the enrichment of the ICM follows the same functional form as the cumulative star formation in field galaxies \cite{Madau1403.0007}, i.e.\ metals are produced (and immediately dispersed) in cluster environments at the same rate, per unit mass, as in the field. An additional free parameter then allows the enrichment of cluster environments to be slowed or sped up relative to the field; slowing could result from the AGN feedback being delayed relative to star formation, while speedups may follow if star formation in the densest environments proceeds earlier than in the field. Although current data already disfavor significantly delayed enrichment in cluster environments, they remain consistent with arbitrarily rapid enrichment at early times, including models with approximately constant metallicity at all observable redshifts. In contrast, the addition of AXIS measurements will provide tight constraints, at the $<10\%$ level, on the rate of enrichment at early times, and a new anchor for studies of star formation and metal enrichment at high redshifts.   
%improve the constraints on parametrized models of metallicity evolution by a factor of 4-5 (Fig. \ref{fig:metal_evo}), enabling the first clear discrimination between competing models.
%AXIS will also observe directly the periods of most intense metal enrichment in individual systems, when star formation activity and AGN feedback peaked, providing our first window onto the system-to-system scatter in the evolutionary history of these then-young but already titanic structures. 

\section{The circumgalactic medium and connections to the cosmic web}
\subsection{Connections to the cosmic web}
\vspace{-0.5\baselineskip}
\noindent (Contributed by S.A. Walker and K.-W. Wong)
\vspace{0.5\baselineskip}
\label{sec:connections}

\begin{figure}[t]
\hbox{
\begin{minipage}{0.51\textwidth}
\includegraphics[width=1.0\textwidth]{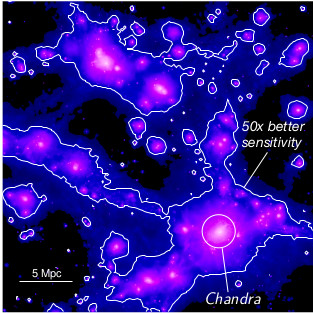}
\end{minipage}
\hfill
\hspace{0.5cm}
\begin{minipage}{0.40\textwidth}
\includegraphics[width=1.0\textwidth]{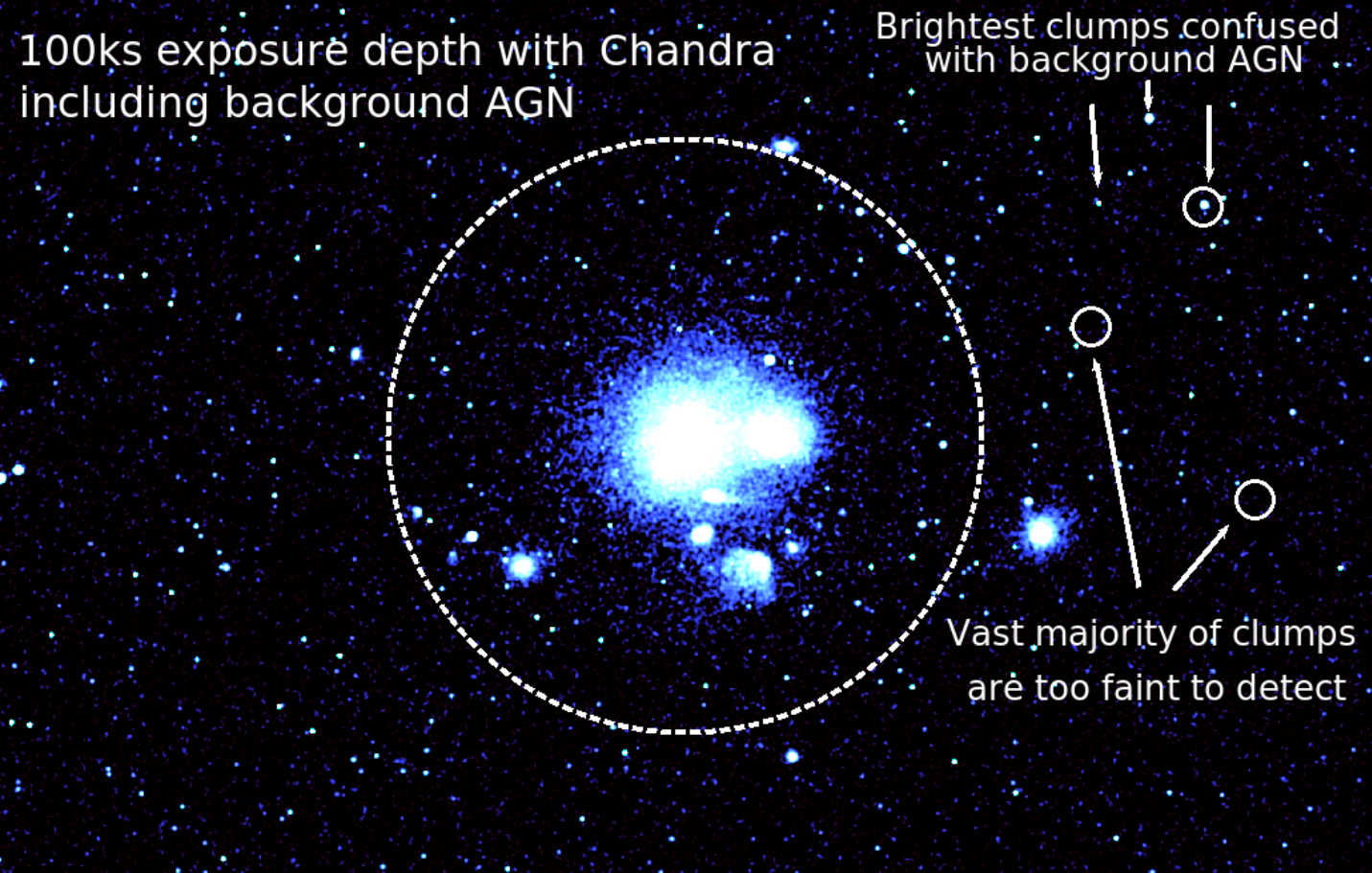}
\includegraphics[width=1.0\textwidth]{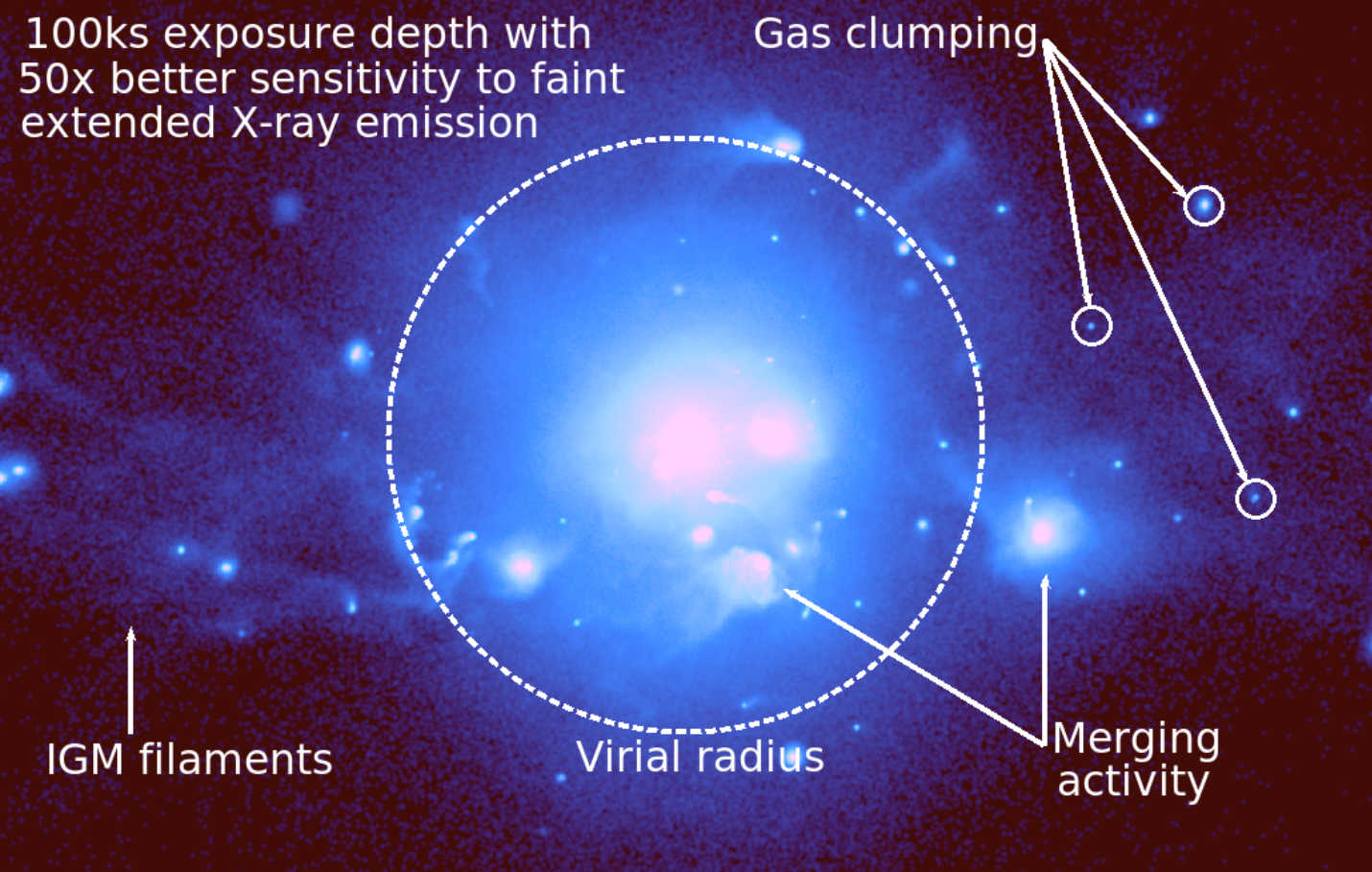}
\end{minipage}
}
\caption[AXIS observations of the outskirts of clusters and the cosmic web]{Left: Hydrodynamical cosmological simulation of large-scale structure formation, showing that galaxy clusters reside at the nodes of IGM filaments \cite{Dolag2006}. With \textit{Chandra} and \textit{XMM-Newton}, only the central regions of clusters can be explored in detail (inner white circle). To begin exploring large-scale structure in its entirety (outer white contours) requires at least a factor of 50 improvement in sensitivity to low surface brightness extended emission. AXIS will achieve this increase in sensitivity through a combination of a high effective area and a low and stable particle background. Right: Simulated X-ray mosaic of the RomulusC cluster \cite{Tremmel2019}, a low mass (10$^{14}$ M$_{\odot}$) cluster at $z=0.05$, comparing 100\,ks deep coverage from \textit{Chandra} (top) and AXIS (bottom). The mock \textit{Chandra} simulation includes background point sources from the cosmic X-ray background. Due to the high background and small collecting areas of current X-ray telescopes, we are only able to see the brightest X-ray emission in the cluster cores. The bulk of the ICM in the outskirts beyond the virial radius (dashed white circle), and the IGM filaments that connect clusters together remain hidden from view. Figure adapted from \cite{Walker_19_whitepaper}.} 
\label{LSS}
\end{figure}

 \noindent The cosmic web forms the backbone of gas flows on cosmological scales. In the local universe, around half of all baryons are expected to lie in the Warm-Hot Intergalactic Medium (WHIM), $10^{5}<T<10^{7}\K$. The hottest part of the WHIM, ($10^{6}<T<10^{7}\K$), can be revealed in the soft X-ray band [see \citealp{Reiprich2013,Walker2022} for a review]. However, due to its very low density, it is very faint in X-ray emission (see Fig. \ref{LSS}). Only with a combination of a large effective area and low background can we observe the cosmic web in emission in X-rays. By combining with X-ray and UV absorption line studies from future missions, AXIS will enable a complete census of the baryons and metals in the local universe for the first time \cite{Richter08}. Cosmological simulations provide predictions for the distribution of metals in the cosmic web and how this distribution depends on AGN feedback over time. AXIS will test these models by charting the distribution of metals in the cluster outskirts and in the WHIM filaments. AXIS will also detect the missing baryons in the CGM around galaxies.

  By measuring gas densities and temperatures in the outskirts of galaxy clusters, we will be able to measure the hydrostatic masses of clusters. Simulations predict that the contribution from non-thermal pressure support should increase in the outskirts, reaching a level of $10\%-30\%$ of the total pressure, and by providing extra support against gravity. Comparing hydrostatic masses with gravitational lensing masses from Euclid and Roman will allow powerful constraints on the levels of non-thermal pressure support in the outskirts of clusters, where they interface with the cosmic web [for a review see \citealp{Walker2019}]. 
  
  AXIS's high spatial resolution, along with its large effective area and low background compared to \textit{Chandra}, would allow it to directly resolve and remove small-scale gas clumps and measure diffuse emission out to $\sim$$2r_{200}$ (corresponding to around 2 Mpc for low mass clusters and 5 Mpc for the most massive clusters).  Gas clumping is believed to be due to infalling gas substructures.  Simulations indicate that the level of gas clumping should increase dramatically in the (1--2)$r_{200}$ region of clusters, where the infalling clumps have not yet been entirely ram-pressure stripped by the ICM \citep{Nagai11_1,roncarelli13,vazza13}. If not resolved, clumping will bias gas profile measurements by overestimating the gas density and the gas mass fraction and underestimating the gas temperature, leading to biases in hydrostatic mass estimates \citep{Simionescu2011,Towler2023}. Detecting and resolving gas clumps, which are associated with infalling galaxies in the outskirts of clusters, is also crucial to gain insight into the stripping of their CGM through their interactions with the ICM \citep{Cen2014}. Removing gas clumps would be critical in order to provide the strongest constraints on electron-ion equilibration and non-equilibrium ionization models in the outskirts of clusters [see section \ref{sec:microphysics} and e.g., \citealp{Wong2009,Wong2011,Akamatsu2011,Avestruz2015,Andreon2023}].  Current X-ray telescopes are unable to resolve faint gas clumps that contribute to the bulk of the clumping (Fig. \ref{LSS}, right), and their physical properties are almost completely unexplored.

\subsection{Inflows and outflows from galaxy clusters}
\vspace{-0.5\baselineskip}
\noindent (Contributed by C. Zhang)
\vspace{0.5\baselineskip}

\noindent Galaxy clusters are assembled through smooth accretion and occasional mergers of small structures (e.g., galaxies, groups) from the cosmic web \citep[see] [for reviews]{Kravtsov2012,Vikhlinin2014}. The newly-accreted IGM (a.k.a., inflows) is heated by accretion shocks at the boundary of the ICM and piles up in the cluster outskirts \citep{Bertschinger1985}. Mergers, on the other hand, not only stir cluster cores, but also drive outflows that can significantly expand the hot atmosphere \citep{CZhang2020a}. It is crucial to study these inflow and outflow processes, not only to gain a deeper understanding of the hierarchical structure formation theory but also to leverage galaxy clusters as powerful tools to investigate cosmology and plasma physics.

\begin{figure*}[!t]
\begin{minipage}{\textwidth}
\centering
\includegraphics[width=0.9\linewidth]{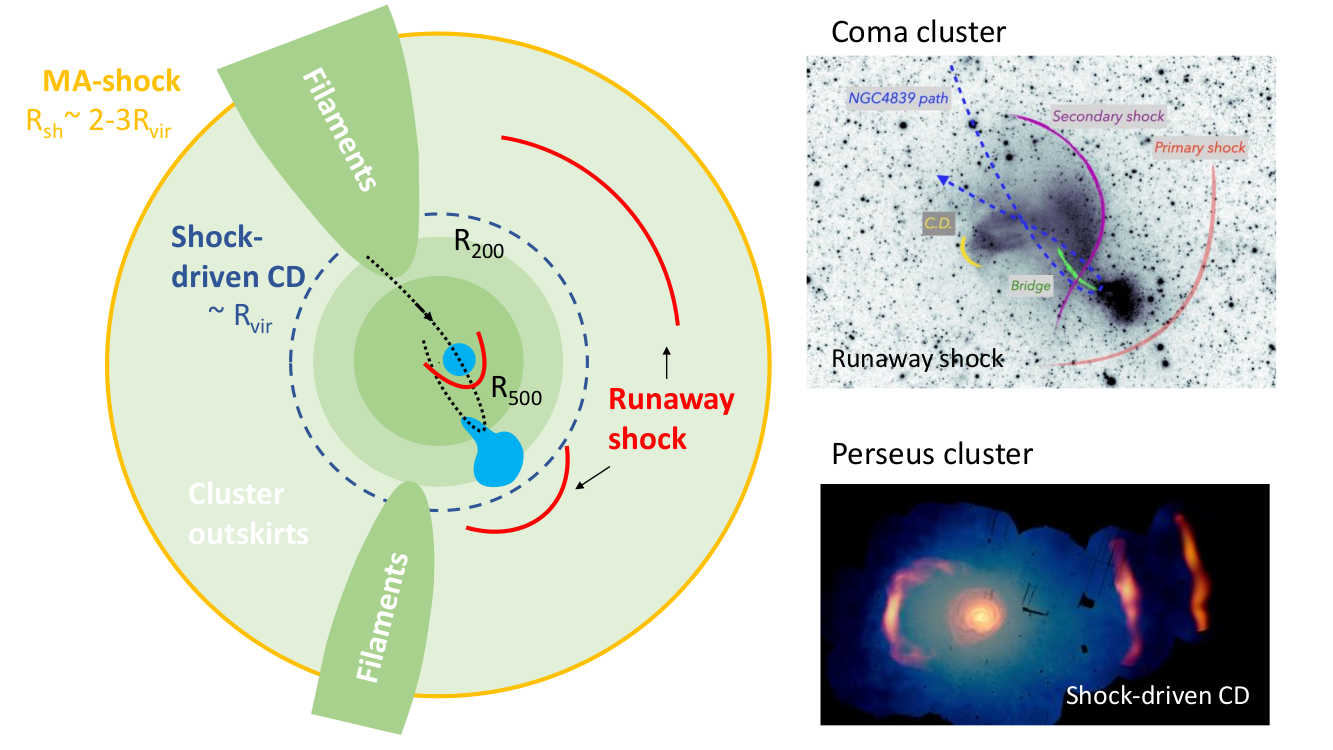}
\caption[Structure in the outskirts of clusters]{A sketch summarizing how mergers and smooth accretion shape the outskirts of galaxy clusters [see also Fig.~1 in \citealp{CZhang2020b}]. Three green circles depict matter distributions within the cluster radius $R_{500}$, $R_{200}$, and external shock radius $R_{\rm sh}$, respectively. An infalling subcluster (blue) moves in the main cluster along the trajectory shown as the dotted black line. The red arcs illustrate merger shocks, which experience a transition from bow shocks to runaway shocks \citep{CZhang2019}. The latter associate with radio relics in the cluster peripheries. The Coma cluster provides a beautiful example supporting this scenario [figure adapted from \citealp{Churazov2021}; see also \citealp{Lyskova2019}]. The long-lived runaway shocks eventually overtake the accretion shock and shape the new boundary of the ICM (i.e., a merger-accelerated accretion shock or MA-shock). In this process, Mpc-scale contact discontinuities (CD) are formed near $R_{\rm vir}$ [blue dashed line in the sketch; see \citealp{CZhang2020b}]. One such candidate has been tentatively identified in the Perseus cluster [Gaussian gradient magnitude filtered map in red over the \textit{XMM-Newton} mosaic in blue; figure adapted from \citealp{Walker2022CF}]. AXIS will dramatically advance our ability to measure sharp gaseous structures (e.g., runaway shocks, contact discontinuities) in the cluster outskirts, essential to understand the assembly history of galaxy clusters.}
\label{fig:cluster_outflow}
\end{minipage}
\end{figure*}

Sharp gaseous structures, including shocks and contact discontinuities [for a review, see, e.g., \citealp{Markevitch2007}], provide rich information on inflows and outflows of the ICM, e.g., their formation mechanisms, energy transportation, and dissipation. Current X-ray detections, performed by, e.g., \textit{Chandra} and \textit{XMM-Newton}, are typically limited to within $R_{500}$. 
Fig.~\ref{fig:cluster_outflow} illustrates a general picture of gaseous structures in the cluster outskirts, highlighting the significance of resolving discontinuities outside $R_{500}$ to understand the dynamics of galaxy clusters. Bow shocks formed ahead of infalling subhalos eventually detach from subhalos themselves after the primary core passages, occurring typically between $R_{500}$ and $R_{200}$, and propagate all the way to the cluster peripheries \citep{CZhang2019,CZhang2021a}. The collisionless nature of these ``runaway’’ shocks makes them ideal targets to characterize the electron-ion non-equilibrium and constrain the plasma physics of the ICM. Given the typical merger rate of galaxy clusters \citep[e.g.,][]{Fakhouri2010} and a combination of high spatial resolution and large effective area, AXIS is expected to detect a large number of such shocks. Together with radio surveys, we will be able to connect X-ray shocks with large-scale radio relics and understand how shocks with moderate Mach numbers accelerate electrons \citep[see][for reviews]{Bykov2019,vanWeeren2019}.

The evolution of runaway shocks largely depends on the steepness of the gas density radial profiles for the diffuse ICM \citep{CZhang2019}. AXIS will provide an unprecedented measurement out to $2R_{200}$ by carefully excluding contributions from cold gas clumps (e.g., Fig. \ref{LSS}). This is essential to understand the evolution of merger-driven outflows, e.g., fate of runaway shocks, and how dramatically they can perturb/re-distribute matter and energy in the cluster outskirts. 

Current theoretical studies predict that runaway shocks can be long-lived and eventually overtake accretion shocks and shape a new boundary of the ICM [i.e. merger-accelerated accretion shock fronts or MA-shocks; see \citealp{CZhang2020a,CZhang2021b}], leading to a significant radial offset between the ICM and dark-matter halo boundaries. This is one of the most prominent effects caused by merger-driven outflows. Cosmological simulations show that the boundaries of the ICM (e.g., accretion shock fronts) are up to $5-6R_{200}$ in the non-filamentary directions \citep[e.g.,][]{Lau2015,Aung2021}. Detecting the Mpc-scale contact discontinuity near the virial radius formed together with the MA-shock can provide a direct evidence of the merger-driven outflow and how it breaks ICM self-similarity in the cluster outskirts \citep{CZhang2020b}. One such candidate has been tentatively identified in the Perseus cluster from deep \textit{XMM-Newton} observations [see Fig.~\ref{fig:cluster_outflow} and also \citealp{Walker2022CF}]. AXIS will not only provide detailed structures of the discontinuity interfaces (e.g., instabilities, mixing), but also extend the sample size, allowing a systematic study of the gas dynamics in the cluster outskirts.  

Last but not least, along filamentary directions, gas dynamics can be very different from the other diffuse directions, due to the anisotropy of the mass accretion rate \citep{CZhang2021b,Vurm2023}. Deeply-penetrated filaments in the intracluster medium have been measured in cosmological simulations \citep[e.g.,][]{Zinger2016} and result in accretion shocks near and even within $R_{200}$. AXIS will be able to detect these innermost accretion shocks and characterize their role in thermalizing IGM and quenching infalling galaxies along the filaments.

\begin{figure*}
\begin{minipage}{\textwidth}
    \centering
    \raisebox{0.5cm}{\includegraphics[width=0.4\textwidth]{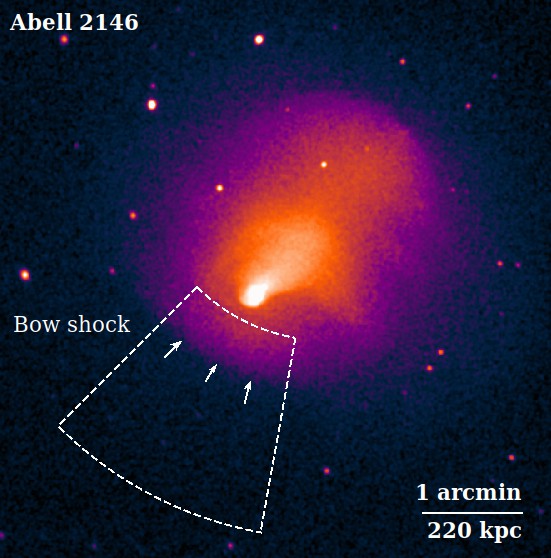}}
    \includegraphics[width=0.55\textwidth]{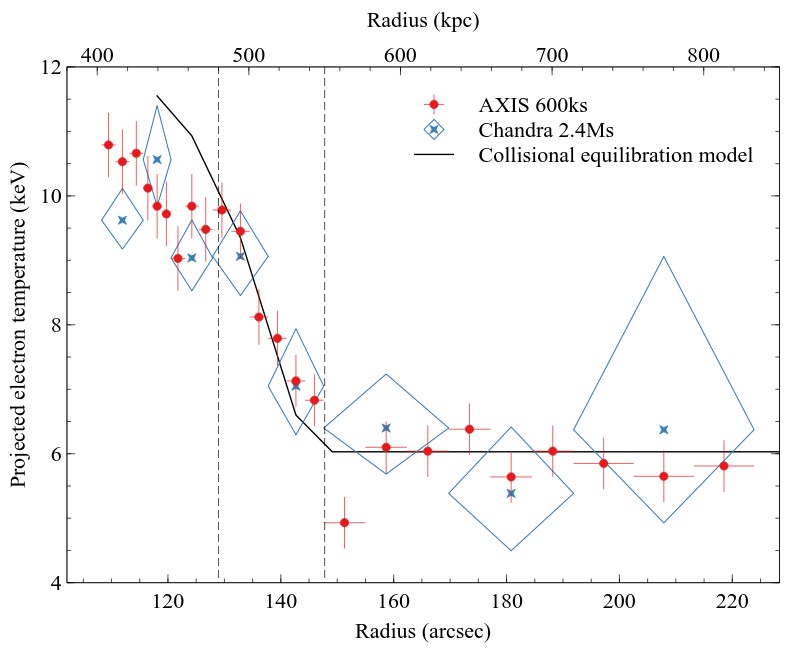}
    \caption[Electron-ion equilibration behind shock fronts with AXIS]{Left: \textit{Chandra} image of the merging galaxy cluster Abell\,2146 \cite{Russell22}.  Right: Projected electron temperature profiles extracted for the sector across the bow shock front at a radius of $\sim150\asec$ (shown on the left).  The electrons are not directly heated by the shock front and instead come back into thermal equilibrium with the shock-heated ions behind the shock.  The expectation for the collisional heating timescale is shown by the blue line.  However, systematic uncertainties increasingly dominate with distance behind the shock (shown by the low temperatures at radii $<130\asec$ compared to the collisional model).  AXIS will resolve the critical region of the postshock gas (between the dashed lines) where equilibration models can be effectively distinguished.}
    \label{fig:microphysics}
    \end{minipage}
\end{figure*}

\subsection{Microphysics of the intracluster and circumgalactic medium}
\vspace{-0.5\baselineskip}
\noindent (Contributed by P.P. Choudhury and C.S. Reynolds)
\vspace{0.5\baselineskip}
\label{sec:microphysics}

\noindent The ICM and CGM are massive reservoirs of dilute, optically thin, weakly collisional (electrons and ions not necessarily in thermal equilibrium by Coulomb collisions), magnetized (with thermal to magnetic pressure $>100$) and hot ($\sim 10^5-10^7\K$) plasmas which are ideal testbeds for microphysical transport processes [e.g., \citealp{helander2005collisional, 2009ApJS_scheko}]. Since AXIS will allow the census and mapping of hot plasma out to the cluster/halo virial radius (see section \ref{sec:connections}), energy transport across large volumes of these astrophysical environments can be studied. Cosmological hydrodynamic galaxy formation simulations apply phenomenological ``sub-grid" prescriptions [e.g., \citealp{2013MNRAS_vogelsberger, 2015MNRAS_crain}] for AGN feedback at galactic scale. These models are not informed by the underlying physical transport and thermalization mechanisms. Thus, there is no good way to constrain how the diffuse plasma is heated and maintained despite an overall consensus on the source of energy. Moreover, this limits our understanding of how much multiphase gas forms in the ICM/CGM - a key unknown of AGN feedback models (sections \ref{sec:radiomodegals} and \ref{sec:clusterfeedback}). Although the magnetic field permeating the diffuse plasma is weak, its dynamical role is highlighted in fully collisionless models of the plasma [e.g., \citealp{2016MNRAS_komarov}] and more recently in magnetohydrodynamics \cite{2021ApJ_drake}. In the presence of typical magnetic fields and large-scale gradients of macroscopic variables (like density, temperature, etc.), a host of microinstabilities are generated at the electron/ion Larmor scale. These instabilities (e.g., whistlers/mirror/firehose) can grow to large amplitudes and scatter off electrons in the nonlinear stage to alter transport properties in weakly magnetized plasma [e.g., \citealp{PhysRevLett2008_scheko}]. While the impact of this phenomenon at global scale is not widely explored [but see \citealp{2021MNRAS_berlok, 2022A&A_beckmann}] even in theoretical/computational research due to the large separation of scales, the consequences for AGN energy transport can be profound. The other aspect of energetics that is not explored sufficiently is the dissipation mechanism (kinetic to thermal energy conversion). Dissipation is crucial to combat the rapid cooling via free-free emission, especially in clusters. Overall, it is invaluable to experimentally measure these key plasma properties like electron-ion equilibration, energy transport, dissipation, role of magnetic field, etc to prescribe accurate ``sub-grid" computational models. 

\begin{figure*}
\begin{minipage}{\textwidth}
    \centering
    \includegraphics[width=0.9\textwidth]{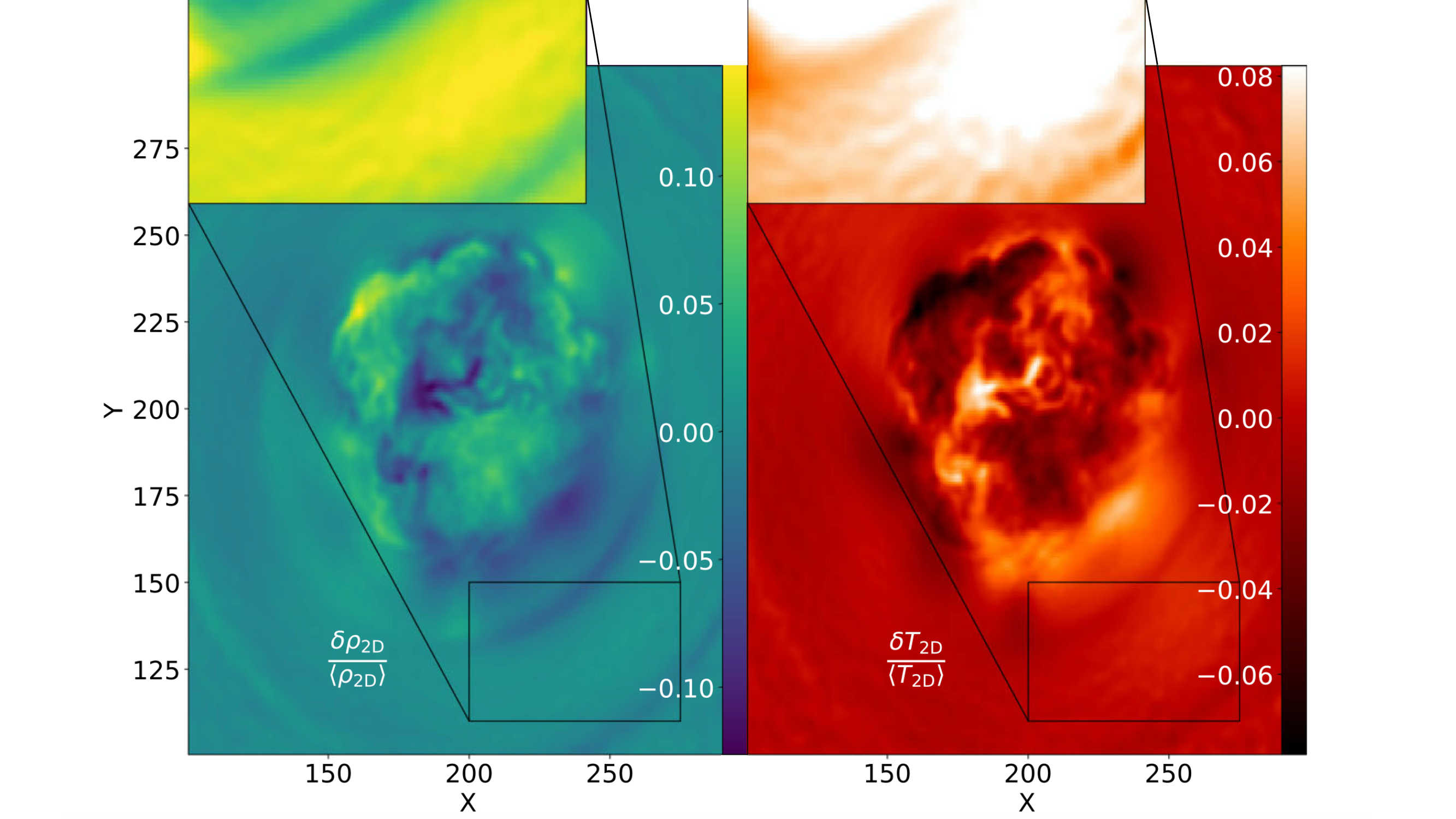}
    \caption[Fluctuations in emission-weighted density and temperature in a 3D hydrodynamic simulation of galaxy cluster]{Fluctuations in emission-weighted density (left) and temperature (right) on a given plane in a 3D hydrodynamic simulation of galaxy cluster \cite{2022MNRAS_choudhury}. The central turbulent region is volume-filling and contributes dominantly to isobaric fluctuations. The correlation of the compressible fluctuations are highlighted in the zoom-in. If transport and dissipation are inefficient in cluster plasma, these features can be sharp. AXIS can verify the presence/absence of such spatially concentrated compressible features. This simulation incorporates idealized isotropic AGN feedback with injection frequency motivated by bubble dynamics in clusters.}
    \label{fig:residuals}
    \end{minipage}
\end{figure*}

AXIS will be instrumental in analysing shock discontinuity and contact discontinuity [in terms of electron temperature and density; see \citealp{MARKEVITCH20071}] to  assess the impact of plasma physics at macroscopic scales. First, the ratio of electron to ion temperature [e.g., \citealp{2009ApJ_wong}] behind the shock front is dependent on the Coulomb collisional timescale and plasma turbulence (which can transfer energy between the species). Very deep \textit{Chandra} observations of a few of the brightest merger shocks have not produced a consensus on the equilibration timescale and mechanism from the electron temperature structure [Fig. \ref{fig:microphysics}, see \citealp{Markevitch06,Wang18,Russell22}]. AXIS's large effective area will allow us to map the detailed temperature structure in the key region close to the shock front where equilibration models can be effectively distinguished. Secondly, the contact discontinuity (``cold front") will be smeared if the Kelvin-Helmholtz instability is strong at the interface (not viscous). Weakly compressible wave features will be sharp if energy transport and viscous dissipation processes are inefficient in cluster plasma (Fig. \ref{fig:residuals}). Moreover, an assessment of the survival timescale of jet-inflated bubbles can probe the levels of viscosity. Third, the electron temperature gradient maps can provide information on electron-dominated conduction [\citealp{2000MNRAS_ettori}], especially across shock fronts. Smaller gradients and/or an electron temperature precursor (electron temperature gradually increases from pre-shock to post-shock region) near a shock discontinuity will imply efficient transport. These investigations can be enabled via AXIS's combination of high spatial resolution and large effective area to obtain high signal-to-noise data on the scale of the electron mean-free path. 

\section{Formation and evolution of large-scale structure}
\subsection{IGM heating by early X-ray sources}
\vspace{-0.5\baselineskip}
\noindent (Contributed by K. Garofali and B.~D Lehmer)
\vspace{0.5\baselineskip}

\noindent Current and upcoming facilities are primed to peer back to the earliest cosmic epochs. JWST is now observing galaxies during the epoch of reionization (EoR), and $21\cm$ interferometers such as the Hydrogen Epoch of Reionziation (HERA) array will soon be sensitive to signals as far back as $z \sim 27$ \citep{HERAPhaseI}. The cosmic $21\cm$ signal is a window into the ionization state and thermal history of the intergalactic medium (IGM), which is sensitive to the high-energy (UV--to--X-ray) photons from the first galaxies \citep[e.g.,][]{McQuinn2016}. In particular, the X-ray emission from the first galaxies can both ionize and heat the neutral IGM, given X-ray photons' longer mean free path. However, the timing of X-ray heating in the early Universe depends critically on assumptions about the sources of high-energy photons in early galaxies \citep[e.g.,][]{Mesinger2013}. 

Recent theoretical models \citep[e.g.,][]{Fragos2013,Madau2017} predict that compact luminous accreting {\it stellar mass} objects (i.e. X-ray binaries, XRBs) would have dominated the X-ray emissivity at redshifts $z > 5$, leading to an epoch of heating (EoH) that precedes the EoR. However, predictions for X-ray heating at $z > 10$ from star-forming galaxies are based on extrapolations of X-ray emissivity trends that are constrained only at $z < 3$, a regime where galaxies with actively accreting central engines dominate X-ray emissivity.
%power output measured from relatively local star-forming galaxies (i.e., $z < 3$), corresponding to the redshift range where galaxies with actively accreting central engines dominate the X-ray emissivity. 
For XRB emission from star-forming galaxies, the deepest existing survey fields yield direct detections in the hundreds of galaxies out to $z \sim 1$, with further measurements of XRB radiative power only out to $z \sim 3$ using high signal-to-noise stacks \citep{Lehmer2016}. Therefore, we currently lack direct measurements of when XRBs begin to dominate the X-ray emissivity, limiting our understanding of the role of XRB feedback during key epochs in cosmic history, such as the EoH. AXIS, with its broad bandpass, high spatial resolution that is well matched to current and future facilities at other wavelengths, stable PSF across the FOV, and high effective area, will directly measure the X-ray emission from star-forming galaxies out to $z \sim 6-8$. With the nominal AXIS deep survey field, detections of XRB emission from star-forming galaxies will number in the thousands per redshift bin, providing a direct measurement of the redshift evolution of XRB emissivity. In this way, AXIS will bridge the gap between locally measured
XRB power output and upcoming indirect measurements of X-ray emission from the first galaxies at $z > 10$ from $21\cm$ interferometers.

\begin{comment}
\subsection{First groups and clusters}
\vspace{-0.5\baselineskip}
\noindent (Contributed by S.W. Allen and M. McDonald?)
\vspace{0.5\baselineskip}

\noindent {\it We still need the simulations that show this but ...}

\noindent Observations of distant galaxy groups provide another important window onto the physics of galaxy formation. At high redshifts, the first galaxy groups to form can only be detected reliably by highly sensitive X-ray surveys. With its large collecting area, excellent soft X-ray response and high spatial resolution across a wide field of view (providing the ability to separate extended emission from point-like sources), AXIS will provide an unprecedented tool for such studies. Cosmological simulations predict that the [how many fields?], [how many MS] the AXIS deep survey fields should detect $\sim$ [TO DO] galaxy groups at [TO DO] $z > *$ down to a mass limit of $M> *$ at $z = 4-5$ [check]. Interestingly, this mass scale is comparable to the largest individual galaxies that AXIS will study at low redshift, providing a vast redshift lever arm over which to study the impact of feedback processes. 
\end{comment}

\subsection{Protocluster structure}
\vspace{-0.5\baselineskip}
\noindent (Contributed by P. Tozzi)
\vspace{0.5\baselineskip}

\noindent The large-scale environment is expected to have a major influence on galaxy evolution, particularly in so-called protoclusters \citep{Overzier16}.  These are 
defined as overdense regions on scales of several physical Mpc at $z>2$ and are destined to
evolve into galaxy clusters with masses in the range $10^{14}-10^{15} M_\odot$ 
by $z=0$ \citep{Chiang13}. Gravitational processes, coupled with the larger availability of diffuse gas at
higher densities and a high merger rate
% (e.g., Hine et al. 2016, Oteo et al. 2018), 
are expected to drive enhanced star formation and a higher AGN duty cycle in protocluster galaxies, 
significantly affecting their evolution. % \citep[see, e.g.][]{DeLucia06}.
In addition, according to the hierarchical model of structure formation, a few halos may be
already virialized within the protocluster, heating the diffuse baryons and lighting up the proto-intra-cluster medium (proto-ICM) in emission in the X-ray band. 

% \begin{figure}
%     \centering
%     \includegraphics[width=0.33\columnwidth]{Fig1_left_PTozzi22.png}
%     \includegraphics[width=0.33\columnwidth]{Fig2_center_PTozzi22.png}
%     \includegraphics[width=0.33\columnwidth]{Fig3_right_Lepore23.png}
%     \caption{Left: background-subtracted soft-band image of the Spiderweb Galaxy 
% after AGN subtraction obtained with a 700 ks {\sl Chandra} observation in AO20. 
% The image has been smoothed for clarity.  
% The green and magenta boxes highlight the east and west radio jets, respectively. 
% Red contours show radio emission observed in the 10 GHz band with the JVLA
% (Carilli et al. 2022) at levels of $0.03$, $0.2$, $2$ and $20$ mJy/beam. 
% Center: measurements of the AGN fraction 
% available in the literature as a function of redshift.  The Spiderweb Protocluster
% is marked with a blue dot. Right: temperature profile as a function of the distance from the central radio
% source for different values of the central density (which is unconstrained). 
% The solid black line represent the average temperature value of the ICM found by Tozzi et al. (2022), 
% while the shaded regions mark the uncertainties on the temperature.}
% \label{Spiderweb}
% \end{figure}

Many of these aspects were recently highlighted in a deep \textit{Chandra}
observation of the archetypal Spiderweb protocluster
\citep{Tozzi22a}, where a high AGN fraction of $25\pm 5$\% (corresponding to an enhancement of $\sim 6$ 
with respect to the COSMOS field) has been measured in 
members with ${\log}(M_*)>10.5$. The current measurement of the X-ray AGN fraction  
in protoclusters provides a confusing picture \citep[see][]{Vito20,Polletta21,Tozzi22a}, 
presumably obscured by the diversity of large-scale structure 
investigated so far. With AXIS, we will routinely probe
luminosities below $L_{[2-10\keV]}10^{43}\ergps$ and measure intrinsic absorption, iron line properties, and spectral
slope in most of the protocluster AGN.  AXIS will thereby reveal whether the X-ray nuclear emission 
is triggered by mergers, gas infall, or secular processes, 
% providing a coherent and physically 
% motivated picture of the evolution of the AGN fraction, 
and ultimately constrain the AGN duty cycle as a function of local overdensity, redshift and stellar mass of the host. 
In addition, the interactions of X-ray-emitting AGN and the 
warm diffuse baryons across the entire proto-ICM may explain the
presence of large Ly$\alpha$ emission-line nebulae extending across several hundreds of kpc
that have been
recently discovered at high-z with integral-field unit spectrographs \citep[][]{Cantalupo14, Arrigoni18}.

\begin{figure}
    \centering
    \includegraphics[width=0.49\columnwidth]{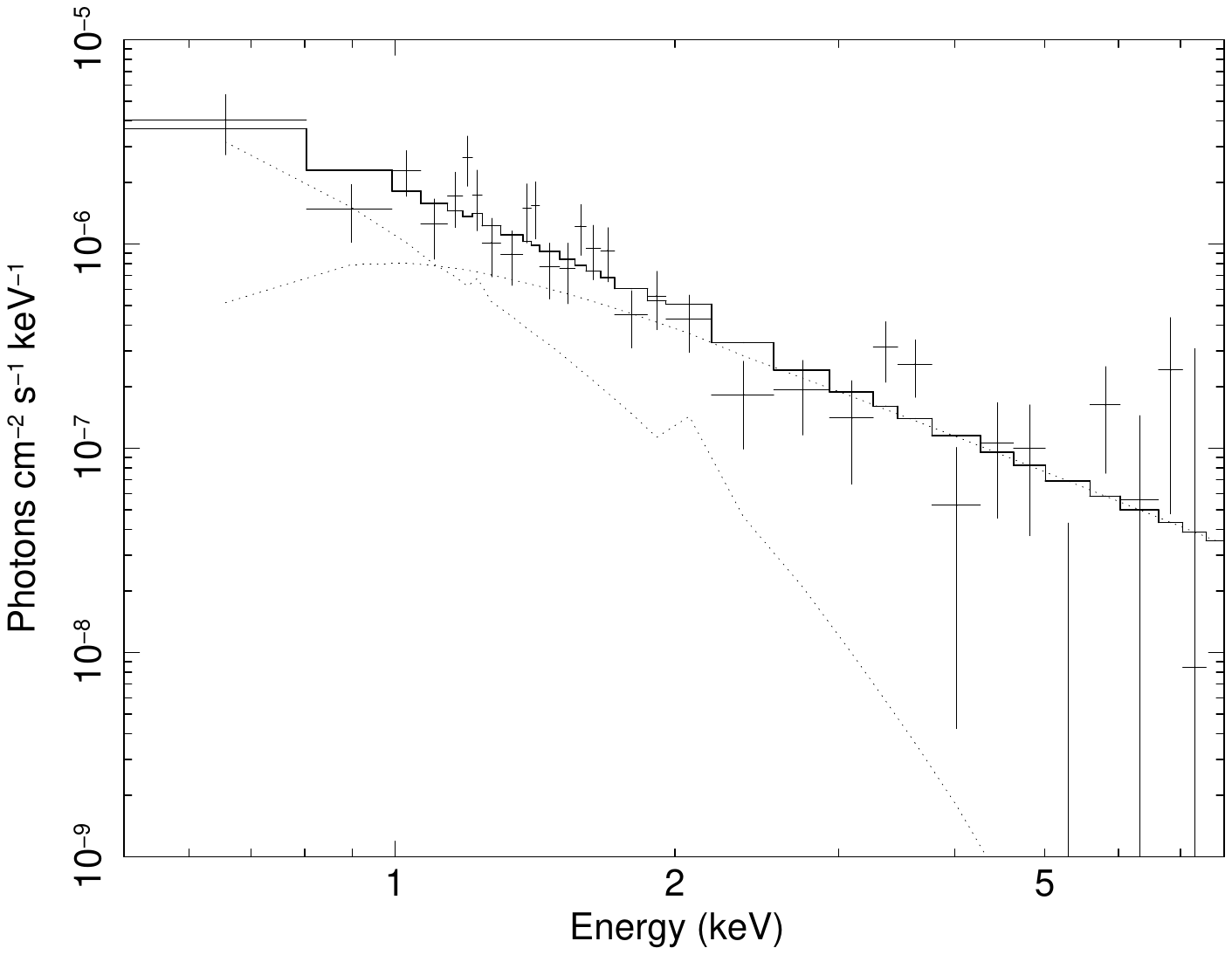}
    \includegraphics[width=0.49\columnwidth]{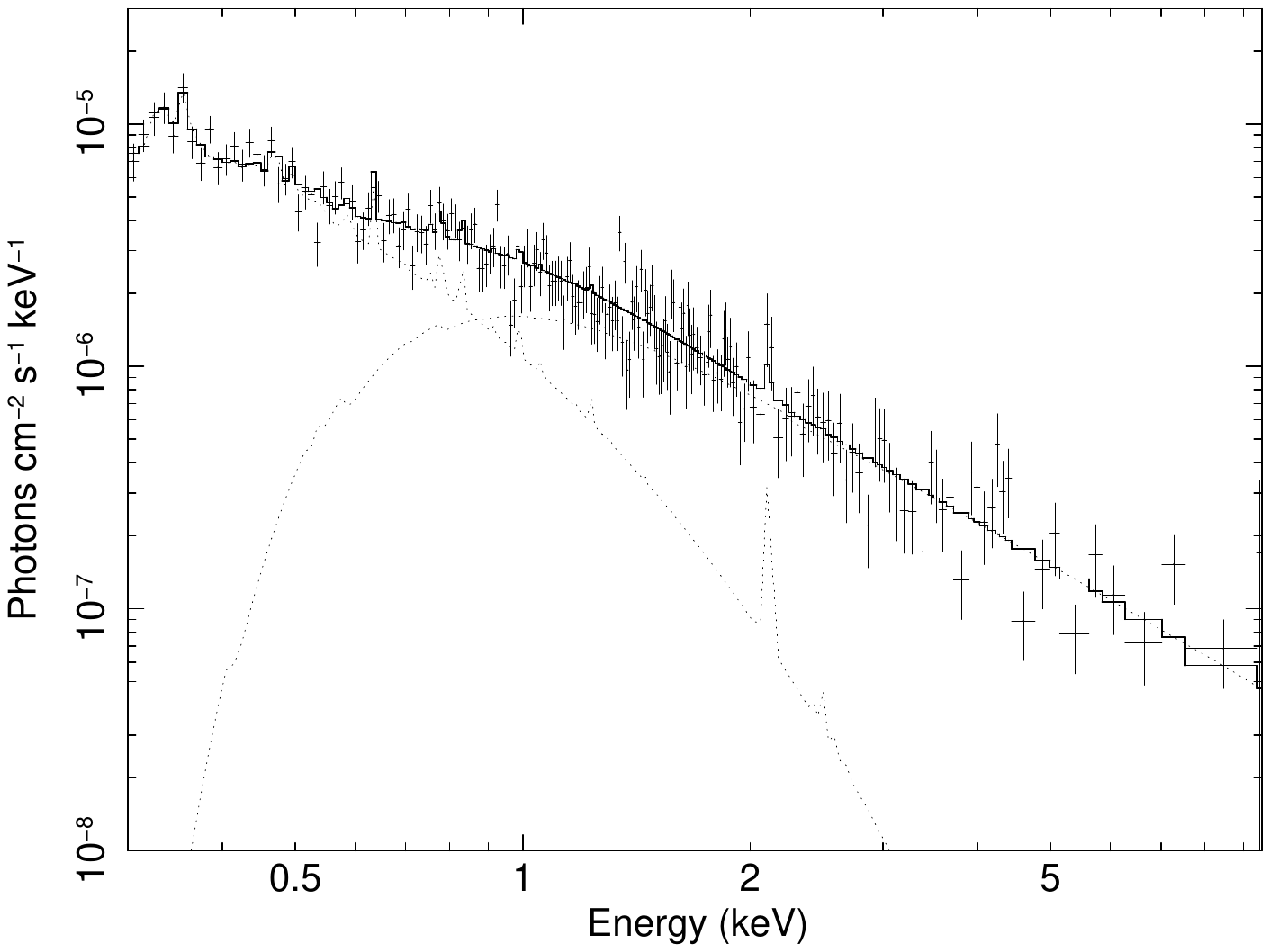}
    \caption[\textit{Chandra} and simulated AXIS spectra of the Spiderweb galaxy]{Left: the \textit{Chandra} spectrum (unfolded) of the thermal emission 
    from the Spiderweb Galaxy halo within $12\asec$ removing the AGN and jets, 
    from the entire exposure of $715\ks$.  Solid lines show the best fit
    obtained with a thermal {\tt mekal} model plus
the AGN contamination. Right: the unfolded spectrum of this target from a simulated $200\ks$ AXIS observation at the aimpoint. To account for the larger
PSF, the AGN contamination is twice that of the \textit{Chandra} spectrum. }
\label{Spiderweb_spectrum}
\end{figure}

% Protoclusters are expected to coincide with intersections
% of dense, gas-rich filaments of the forming cosmic web (e.g., Umehata+19). 
% Observations with narrow-band filters and are revolutionizing the 
% field of diffuse baryons in high-redshift overdense regions, 
% by discovering extending up
% to  (e.g., Cantalupo et al. 2014, Hennawi et al. 2015, Arrigoni-
% Battaia et al. 2018), possibly powered by photoionization from star-forming galaxies and AGN
% in the protocluster. 

Another key scientific topic is the formation and evolution of the proto-ICM. 
Thanks to \textit{Chandra}, a roughly spherical ICM halo with $kT \sim 2\keV$  
has been found around the Spiderweb Galaxy \citep[see Fig. 13 in][]{Tozzi22b}, 
clearly showing that all unresolved and diffuse components can be identified and characterized with $\sim $ arcsec angular resolution, even in the most complex case where thermal and non-thermal 
emission largely overlap due to the presence of a 
bright radio galaxy. 
Coupling high angular resolution to the large effective area of AXIS, 
particularly at soft energies, will revolutionize protocluster studies in every 
aspect. In medium-deep exposures (of the order of $200-300\ks$)
with AXIS, the combination of imaging and spectral analysis will allow us to measure the chemical 
enrichment of the proto-ICM, the presence of temperature gradients, and possibly cooling flows 
(that can host much larger mass deposition rates compared to those at low redshifts), 
and the presence of cavities carved within the proto-ICM by mechanical feedback. 
In Fig. \ref{Spiderweb_spectrum},
we compare the \textit{Chandra} spectrum of the Spiderweb to that expected from a $200\ks$ exposure
with AXIS at the aimpoint, finding that we can 
measure the average proto-ICM temperature with an accuracy of $0.1\keV$ and the proto-ICM chemical abundance
with an uncertainty of 20\% for a reference case with an average temperature below 2 keV and an average metallicity of $0.3Z_\odot$ at $z\sim 2$. 
% This example shows that even in the most complex case, AXIS pave the way to a detailed, 
% spatially resolved analysis of the ICM well beyond $z\sim 2$. We also remark that the sensitivity
% below 0.5 keV strongly improves the accuracy of temperature and abundance measurements. 

% At $z>2$, we do expect that the interactions between the ICM and the forming BCG is so strong 
% that we may reveal with high accuracy several key processes
% that are expected to act with less intensity across the entire life of the cluster.  

Many other fields of investigation are made possible by AXIS.
The spectral characterization of the nonthermal, diffuse X-ray emission from radio jets due to inverse Compton scattering onto the 
photons of the cosmic microwave background (and possibly local IR photons), 
allows us to directly constrain the magnetic fields and pressure around the jets, 
shedding light on the interactions between the relativistic electrons and the diffuse 
ICM \citep[see][]{Carilli22, Anderson22}.  
% AGN feedback is expected to impact the host and the surrounding galaxies in a non trivial way, 
% sometimes quenching and sometimes promoting star formation. A notable
% example is the first evidence for increased star-formation rate due to the effect of a radiogalaxy
% on external galaxies in a z = 1.7 protocluster (Gilli et al. 2019). 
In addition, with AXIS's high spatial resolution, it will be possible to investigate
the X-ray emission from star formation in a single, strong starburst, and 
as an average across protocluster galaxies.  
% The most extremely star-forming protoclusters have been 
% identified as overdensities
% of dusty star-forming galaxies at $z \geq 4$ (e.g., Oteo et al. 2018, Milleret al. 2018, Ivison et al. 2020).
Considering that a star formation rate of 
$100\Msunpyr$ corresponds to $\sim 3\times 10^{-17}\ergpcmsqps$ at $z=2.5$, and that the 
cumulative SFR integrated over the protocluster members can amount to several thousands, the
X-ray emission from star formation can be detected through stacking down to an average level of 
few $\Msunpyr$ over a population of $\sim 100$ protocluster members. 
In this way, AXIS will map the SFR from the central regions to
the outskirts and constrain the triggering mechanisms 
of star formation episodes in protoclusters \citep[see][]{Chiang17}.  
Finally, the hot ICM in virialized halos at temperatures above $1\keV$ can be serendipitously detected
at any distance, considering that the typical surface brightness of virialized halos does not
decrease significantly with redshift \citep{Churazov15}.
For the faintest halos, for which an X-ray 
spectral analysis is not viable, it is possible to recover the temperature and 
the entropy profiles by exploiting the
complementarity of SZ and X-ray observations, as demonstrated in the 
Spiderweb protocluster \citep[][Lepore et al. in preparation]{DiMascolo23}.
 
% To summarize, protoclusters allow
% us to probe the early phases of galaxy and SMBH growth, AGN feedback, and ICM formation,
% that shaped the observed properties of present-day cluster galaxy population. 
% The number of known protoclusters
% is still relatively low, but it has been constantly increasing in the last 10 yrs thanks to the
% applications of several selection techniques, including wide-field spectroscopic and photometric surveys
% (e.g., Cucciati+14, Toshikawa+16), narrow-band imaging (Ouchi+05; Higuchi+19), sub-mm
% observations (e.g., Oteo+18, Hill+20), and the use of beacons 
% such as powerful radio galaxies (e.g., Pentericci+00; Venemans+02; Gilli+19,
% Brienza+23), QSOs up to z = 6.3 (e.g., Mignoli+20), or extended Ly$_\alpha$ emission over tens to-
% hundreds kpc scales (e.g., Cantalupo+14, Arrigoni-Battaia+18). 
The study of the formation, evolution, and physical properties of protoclusters is one of the most
vibrant fields in modern astrophysics and is rapidly advancing with new state-of-the-art instrumentation,
such as JWST, ALMA, and MUSE@VLT. 
% In particular, JWST is proving a truly transformational facility in
% this field by discovering new protoclusters in the infant universe at an impressive rate (Kashino+23,
% Wang+23, Morishita+23, Castellano+23, Helton+23). 
Furthermore, future survey missions, such as Euclid and the Vera Rubin Observatory,
will prioritize the study of protoclusters. In the near future, these
facilities will boost the number of known and well characterized protoclusters, 
providing a large sample of large scale overdensities across different epochs,
with vastly different properties. 

% I think the above sentence is a solid ending for this section

%A facility like AXIS is the only way to trace a wealth %of high-energy phenomena that would be otherwise missed %at other wavelenghts or
%at poor resolution. 

% to provide the  X-ray view of these targets will have an invaluable 
% scientific return in galaxy formation and evolution. 

\section{Conclusions}

\noindent AXIS will be a powerful community-based X-ray observatory for the 2030s.  The unparalleled combination of arcsecond spatial resolution, high throughput, and wide field of view will allow AXIS to capture the complete picture of feedback from the impact zones, where stellar winds, AGN winds, and radio jets collide with the ISM, to the widespread distribution of energy and metal-rich gas across galaxy and cluster halos.  AXIS's low instrumental background and wide field of view will finally permit a dramatic emergence of the cosmic web as a vast network of soft X-ray filaments, reveal the establishment of feedback in clusters at cosmic noon, and show us the structure of the first groups and clusters at high redshift.  AXIS's primary science goals on galaxy evolution and feedback address the science questions asked by the Astro2020 Decadal `Cosmic Ecosystems' priority area and, as such, will have a wide-ranging impact on the whole astronomical community.  From radio jets (ngVLA, SKA) and the multi-phase gas flows driven by feedback in galaxies and clusters (ELT/TMT, JWST, ALMA) to the detection of massive clusters at high redshift (CMB-S4, Simons Observatory, Rubin, Roman and Euclid), AXIS has synergies with essentially every planned observatory for the 2030s.  Companion white papers on topics including stellar physics and exoplanets, compact object populations and supernova remnants, transients and multi-messenger astronomy, and the evolution of AGN demonstrate the breadth of discoveries that the AXIS probe mission will deliver for the entire astronomical community (see the \href{http://axis.astro.umd.edu/}{AXIS website}).

\acknowledgments{We thank everyone who has contributed to the development of the AXIS Probe mission concept.  HRR acknowledges support from an STFC Ernest Rutherford Fellowship and
an Anne McLaren Fellowship from the University of Nottingham. LAL acknowledges support by the Heising-Simons Foundation and the Simons Foundation. This work was performed in part at the Simons Foundation Flatiron Institute's Center for Computational Astrophysics during LAL's time as an IDEA Scholar.}

\section*{References:}
\vspace{-2.5\baselineskip}
\externalbibliography{yes}
\bibliography{references}

%\bibliographystyle{abbrv}
%\bibliography{references}

\end{document}